\documentclass[aps,twocolumn,superscriptaddress,pra,nofootinbib]{revtex4-1}
\usepackage[ocgcolorlinks,colorlinks=true,linkcolor=blue,citecolor=red,linktocpage=true]{hyperref}
\usepackage{graphicx}
\usepackage{amsthm} 
\usepackage{amsbsy} 
\usepackage{dcolumn}   
\usepackage{verbatim}   
\usepackage{color}     	 
\usepackage{amsmath,amsfonts,amssymb}
\usepackage{natbib}
\usepackage{enumerate}   

\usepackage{graphicx,amsmath,amssymb,color,array}
\usepackage[normalem]{ulem}
\usepackage{times}

\usepackage{amsfonts}
\usepackage{latexsym}
\usepackage{amsmath}
\usepackage{amssymb}
\usepackage{amsfonts}
\usepackage{amsthm}
\usepackage{mathrsfs}
\usepackage{natbib}
\usepackage{color,verbatim,graphics}
\usepackage{psfrag}
\DeclareMathAlphabet{\mathrsfs}{U}{rsfs}{m}{n}
\DeclareMathAlphabet{\mathpzc}{OT1}{pzc}{m}{it}
\DeclareMathAlphabet{\matheus}{U}{eus}{m}{n}
\DeclareMathAlphabet{\mathbbold}{U}{bbold}{m}{n}

\newcommand{\ba}{\begin{align}}
\newcommand{\ea}{\end{align}}

\newcommand{\be}{\begin{equation}}
\newcommand{\ee}{\end{equation}}

\newcommand{\bea}{\begin{eqnarray}}
\newcommand{\eea}{\end{eqnarray}}

\newcommand{\ban}{\begin{eqnarray*}}
\newcommand{\ean}{\end{eqnarray*}}
\newcommand{\Tr}{\operatorname{tr}}
\newcommand{\tr}{\operatorname{tr}}

\newcommand{\ket}[1]{\left|#1\right\rangle}
\newcommand{\bra}[1]{\left\langle#1\right|}

\newcommand{\ketbra}[2]{|#1\rangle\langle#2|}
\newcommand{\expect}[1]{\langle#1\rangle}
\newcommand{\assem}[3]{#1_{#2}^{#3}}

\newcommand{\proj}[1]{\left|#1\right\rangle\left\langle #1\right|}

\newcommand{\ie}{{\it{i.e.}~}}
\newcommand{\etal}{{\it{et al.}}}
\newcommand{\LHS}{\mathrm{LHS}}
\newcommand{\SW}{\mathrm{SW}}
\newcommand{\sep}{\mathrm{sep}}
\newcommand{\SEP}{\mathrm{SEP}}
\newcommand{\rA}{\mathrm{A}}
\newcommand{\rB}{\mathrm{B}}
\newcommand{\rC}{\mathrm{C}}
\newcommand{\rE}{\mathrm{E}}
\newcommand{\rT}{\mathrm{T}}
\newcommand{\rO}{\mathrm{O}}

\newcommand{\obs}{\mathrm{obs}}

\newcommand{\SR}{\mathrm{SR}}
\newcommand{\ER}{\mathrm{ER}}
\newcommand{\SQ}{\mathrm{SQ}}
\newcommand{\guess}{\mathrm{g}}
\newcommand{\qu}{\enskip}

\newcommand{\fs}{\mathrm{fs}}
\newcommand{\bs}{\mathrm{bs}}

\begin{document}
\title{Quantum steering: a review with focus on semidefinite programming}

\author{D. Cavalcanti} 
\affiliation{ICFO-Institut de Ciencies Fotoniques, The Barcelona Institute of Science and Technology, 08860 Castelldefels (Barcelona), Spain}
\email{dcavalcanti@gmail.com}

\author{P. Skrzypczyk} 
\affiliation{H. H. Wills Physics Laboratory, University of Bristol, Tyndall Avenue, Bristol, BS8 1TL, United Kingdom}
\email{paul.skrzypczyk@bristol.ac.uk }

\begin{abstract}
Quantum steering refers to the non-classical correlations that can be observed between the outcomes of measurements applied on half of an entangled state and the resulting post-measured states that are left with the other party.  From an operational point of view, a steering test  can be seen as an entanglement test where one of the parties performs uncharacterised measurements. Thus, quantum steering is a form of quantum inseparability that lies in between the well-known notions of Bell nonlocality and entanglement.  Moreover, quantum steering is also related to several asymmetric quantum information protocols where some of the parties are considered untrusted. Because of these facts, quantum steering has received a lot of attention both theoretically and experimentally. The main goal of this review is to give an overview of how to characterise quantum steering through semidefinite programming. This characterisation provides efficient numerical methods to address a number of problems, including steering detection, quantification, and applications. We also give a brief overview of some important results that are not directly related to semidefinite programming. Finally, we make available a collection of semidefinite programming codes that can be used to study the topics discussed in this article.\end{abstract}

\maketitle
\tableofcontents
\section{Introduction}
\label{introduction}

\begin{figure}
\centering
\includegraphics[width=0.8\columnwidth]{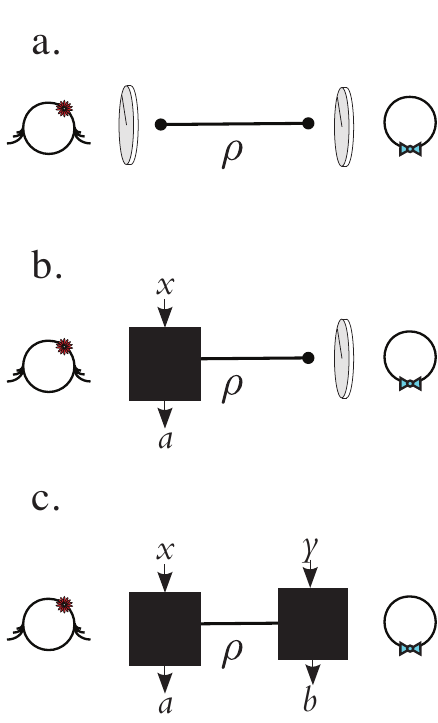}
\caption{Different bipartite scenarios where entanglement can be certified. a. Entanglement scenario: Alice and Bob can apply a tomographically complete set of measurements and retrieve the state $\rho^{\rA\rB}$ they share. The entanglement of $\rho^{\rA\rB}$ can be certified through a sequence of semi-definite programs \cite{DPS02} b. Quantum steering scenario: Alice treats her measurements as a black box with classical inputs $x$ and outputs $a$. Bob performs tomography and reconstructs the set of states $\{\sigma_{a|x}\}$ conditioned to Alice's measurements. In this case, entanglement can be certified through a single SDP test \cite{P13} (see Eq. \eqref{e:feasibility}) c: Quantum non-locality scenario: both Alice and Bob treat their measurements as black boxes. In this case their analysis is based on the set of probability distributions $\{P(ab|xy)\}$ that they can obtain. Entanglement can be detected by running a linear program \cite{BCP+14}}\label{f:scenarios}
\end{figure}

In 1935 E. Schrodinger \cite{S35} introduced the concept of quantum steering in an attempt to formalise the ``spooky action at distance'' discussed by Einstein, Podolsky and Rosen in their seminal paper \cite{EPR35}. Quantum steering refers to the fact that, in a bipartite scenario, one of the parties can change the state of the other distant party by applying local measurements. In 2007 Wiseman, Jones and Doherty formalised steering in terms of the incompatibility of quantum mechanical predictions with a classical-quantum model where pre-determined states are sent to the parties. Furthermore, the observation of quantum steering can also be seen as  the detection of entanglement when one of the parties performs uncharacterised measurements \cite{WJD07}. In this way, steering detection can be seen as a scenario in between a non-locality test \cite{BCP+14} and a standard entanglement test \cite{GT09,HHHH09} (See Fig. \ref{f:scenarios}). Because of this, the study of quantum steering has provided new insights to the understanding of quantum inseparability and consequently this has become a popular area recently. 

The main goal of this article is to give a general overview of quantum steering with a focus on how to characterise it through semidefinite programming. This approach has been recently followed by some researchers and has proved to be useful in several contexts such as steering detection, quantification, and applications. More specifically, we will describe how semidefinite programming can be used to attack the following problems: (i) detect steering and obtain experimental-friendly steering witnesses, (ii) determine the robustness to noise of experimental setups to perform a faithful steering tests, (iii) applications, including randomness certification, entanglement and nonlocality estimation,  (iv) connection to other properties of quantum mechanics such as measurement incompatibility, and (v) foundational problems, such as the characterisation of quantum correlations. 

We stress that the present article is not aimed to be a complete review on quantum steering, but rather provide a general framework to deal with this concept and to provide a set of techniques that can be used in future studies. Nevertheless, we provide in the end of each section a brief overview of some of the most important results on quantum steering.

\section{Definition of quantum steering}
\label{Sec: definition}

We start by defining the scenario in which quantum steering is discussed. Consider a bipartite situation composed by Alice and Bob sharing an unknown quantum state $\rho^{\rA\rB}$. Alice performs $m_\rA$ measurements on her subsystem labelled by $x=0,...,m_\rA-1$, each having $o_\rA$ outcomes $a=0,...,o_\rA-1$. Upon choosing measurement $x$ and receiving outcome $a$, the state of Bob's system is transformed into the state $\rho_{a|x}$ with probability $p(a|x)$. 

The steering scenario consists of the situation where no characterisation of Alice's measurements is assumed, while Bob has full control of his measurements and can thus access the conditional states $\rho_{a|x}$. This is often called a \textit{one-sided device-independent scenario}, as all the results in this scenario do not dependent on any particular information of how Alice's measurements work. The available information in this scenario is the collection of post-measured states and respective conditional probabilities $\{\rho_{a|x},p(a|x)\}_{a,x}$. This information can be summarized by the set of unnormalised quantum sates  $\{\sigma_{a|x}\}_{a,x}$, where $\sigma_{a|x}=p(a|x)\rho_{a|x}$, often called an \emph{assemblage}. According to quantum theory the members of the assemblage can be obtained by
\begin{equation}\label{Born rule}
\sigma_{a|x}=\Tr_\rA[(M_{a|x}\otimes\openone)\rho^{\rA\rB}],
\end{equation}
where $\sum_a M_{a|x}=\openone$ and $M_{a|x}\geq 0 ~ \forall a,x$. We stress that in the steering scenario the state $\rho^{\rA\rB}$ and the measurements $\{M_{a|x}\}_{a,x}$ are completely unknown. Not even Alice's Hilbert space dimension is specified in the problem.

Notice that by definition, quantum assemblages satisfy 
\be\label{nosignaling}
\sum_a \sigma_{a|x}=\sum_a \sigma_{a|x'}=\rho^\rB\quad\forall x,x',
\ee
and
\be\label{normalization}
\tr \sum_a \sigma_{a|x}=1\quad\forall x,
\ee
where $\rho^\rB=\Tr_\rA[\rho^{\rA\rB}]$. Condition \eqref{nosignaling} comes simply from the fact that the average of $\sigma_{a|x}$ over Alice's outcome is the reduced state of Bob (and can be seen as a `no-signalling' requirement from Alice to Bob) while \eqref{normalization} is the normalisation condition.

In the remainder of this article we will denote an assemblage simply as $\sigma_{a|x}$, and a set of measurements as $M_{a|x}$, unless we find it necessary to stress that they actually refer to multiple objects. 

\subsection{Quantum steering as the impossibility of local-hidden-state models}

In 2007 Wiseman, Jones and Doherty formally defined quantum steering as the possibility of remotely generating ensembles that could not be produced by a \textit{local hidden state} (LHS) model \cite{WJD07}. A LHS model refers to the situation where a source sends a classical message $\lambda$ to one of the parties, say Alice, and a corresponding quantum state $\rho_\lambda$ to the other party, Bob. Given that Alice decides to apply measurement $x$, the variable $\lambda$ instructs Alice's measurement device to output the result $a$ with probability $p(a|x,\lambda)$. Additionally it is also considered that the classical message $\lambda$ can be chosen according to a distribution $\mu(\lambda)$. Bob does not have access to the classical variable $\lambda$, so the final assemblage he observes {is composed by the elements
\begin{equation}\label{LHS}
\sigma_{a|x}=\int d\lambda \mu(\lambda)p(a|x,\lambda)\rho_\lambda.
\end{equation}
Notice that the normalisation of each member of the assemblage,  $\tr[\sigma_{a|x}]=\int d\lambda \mu(\lambda)p(a|x,\lambda)=p(a|x)$, gives the probability that Alice observes the result $a$ given that she measured $x$}. In what follows, whenever an assemblage has a LHS decomposition \eqref{LHS} we will denote it by $\sigma_{a|x}^{\LHS}$.

An assemblage is said to demonstrate steering if it does not admit a decomposition of the form \eqref{LHS}. Furthermore, a quantum state $\rho^{\rA\rB}$ is said to be \emph{steerable from $A$ to $B$} ($B$ to $A$) if there exists measurements in Alice's (Bob's) part that produces an assemblage that demonstrates steering. On the contrary, an assemblage is said to be LHS if it can be written as in \eqref{LHS}, and a quantum state is said to be unsteerable if for all local measurements an LHS assemblage is generated\footnote{Note that in the literature one can find the terminology that an assemblage is `steerable' or `unsteerable'. Here, we retain this terminology only for states, and when talking about assemblages use instead `demonstrates steering' or `LHS' as more appropriate descriptions.}. 

It is important to notice that, unlike other notions of quantum correlations like entanglement or non-locality,  the concept of quantum steering is asymmetric: a quantum state can be steerable from Alice to Bob, but unsteerable from Bob to Alice. This phenomenon has been theoretically studied \cite{BVQB14,QVC+15} and experimentally demonstrated recently \cite{HES+12,WWB+16,SYX+16}. 

\subsection{Quantum steering as one-sided device-independent detection of entanglement and measurement incompatibility}
\label{Subsec: SDI ent}

As we will show, by observing steering one can certify, at the same time, that Alice and Bob share entanglement and that Alice's measurements, although uncharacterised, are incompatible. To see how, let us assume that Alice and Bob share a separable state $\rho^{\sep} = \int d\lambda \mu(\lambda) \rho_\lambda^\rA \otimes\rho_\lambda^\rB$. This imposes the following structure on the assemblages created:
\begin{eqnarray}
\sigma_{a|x}&=&\tr_\rA[(M_{a|x}\otimes\openone)\rho^{\sep}]\nonumber\\
&=& \int d\lambda \mu(\lambda) \tr[M_{a|x}\rho_\lambda^\rA]\rho_\lambda^\rB,\nonumber\\
&=& \int d\lambda \mu(\lambda)p(a|x,\lambda)\rho_\lambda^\rB,
\end{eqnarray}
which has the same structure as \eqref{LHS}.

Let us now assume that Alice's measurements are compatible, in the sense of being jointly measurable \cite{BLM96}. This means that there exists a single `parent' measurement $N$ with POVM elements $\{N_\lambda\}_\lambda$, and a collection of probability distributions $p(a|x,\lambda)$, such that {$M_{a|x} = \int d\lambda p(a|x,\lambda) N_\lambda$}. If Alice performs such measurements on an arbitrary state $\rho^{\rA\rB}$, this imposes the following structure on the assemblages created:
{\begin{align}
\sigma_{a|x}&=\tr_\rA[(M_{a|x}\otimes\openone)\rho^{\rA\rB}]\nonumber\\
&= \int d\lambda p(a|x,\lambda)\tr_\rA[(N_\lambda \otimes \openone)\rho_{\rA\rB}],\nonumber\\
&= \int d\lambda \mu(\lambda)p(a|x,\lambda)\rho_\lambda^\rB,
\end{align}
where {$\mu(\lambda) :=\tr[N_\lambda\otimes\openone \rho^{\rA\rB}]$} and $\rho_\lambda^\rB := \tr_\rA[(N_\lambda \otimes \openone)\rho^{\rA\rB}]/ \tr[(N_\lambda\otimes\openone) \rho^{\rA\rB}]$}. This again has the same structure as \eqref{LHS}.

\section{Detection of quantum steering}
\label{Sec: detecting}

As discussed before, deciding whether an assemblage $\sigma_{a|x}$ demonstrates steering amounts to checking whether there exists a collection of quantum states $\rho_\lambda$ and probability distributions $\mu(\lambda)$ and $p(a|x,\lambda)$ such that \eqref{LHS} holds. This is in principle a hard problem, since the variable $\lambda$ could assume infinitely many values. In what follows we show that if the number of measurements and outputs is finite this problem becomes much simpler, and can actually be computed through semidefinite programming (SDP) \cite{VB96}. A brief review of the relevant concepts of SDP can be found in Appendix~\ref{SDP}.

\subsection{The membership problem}

Suppose that $x=0,...,m_\rA-1$ and $a=1,...,o_\rA-1$, i.e. Alice performs $m_\rA$ measurements with $o_\rA$ outcomes each. {A crucial observation is that, given the finite number of measurement choices and outcomes, $p(a|x,\lambda)$ in Eq.~\eqref{LHS} can be decomposed as a convex combination of (a finite number of) deterministic probability distributions. In particular, a deterministic probability distribution $D(a|x,\lambda')$ gives a fixed outcome $a$ for each measurement, \ie $D(a|x,\lambda') = \delta_{a,\lambda'(x)}$, such that $a = \lambda'(x)$, with $\lambda'(\cdot)$ being a function from $\{0,\ldots,m_\rA-1\}$ to $\{0,\ldots, o_\rA-1\}$. One can identify each $\lambda'$ with a string of outcomes $\lambda' = (a_{x=0},a_{x=1},\ldots,a_{x=m_\rA-1})$, such that $\lambda'(x') = a_{x=x'}$. There are $d=o_\rA^{m_\rA}$ such strings, and hence $d$ unique deterministic probability distributions. With this in place, one can always write
\begin{equation}\label{deterministic}
p(a|x,\lambda)=\sum_{\lambda'=1}^d p(\lambda'|\lambda)D(a|x,\lambda'),
\end{equation}
where $p(\lambda'|\lambda)$ is the weight of the deterministic distribution
labelled $\lambda'$ (when the hidden variable takes value $\lambda$). }

By inserting \eqref{deterministic} into \eqref{LHS} and defining {$\sigma_{\lambda'}:= \int d\lambda \mu(\lambda)p(\lambda'|\lambda) \rho_\lambda$} we obtain that an LHS assemblage can be written as
\begin{equation}\label{deterministicLHS}
\sigma_{a|x}=\sum_{\lambda'=1}^{d}D(a|x,\lambda')\sigma_{\lambda'}.
\end{equation}
{Crucially, in comparison to \eqref{LHS}, Eq.~\eqref{deterministicLHS} is a finite sum (instead of an integral), and for each $\lambda'$, $D(a|x,\lambda')$ is a fixed distribution (unlike $p(a|x,\lambda)$, which was unknown). This thus represents a significant simplification in the structure of LHS assemblages. }

Finally, notice that
{
\begin{align}\label{e:normalisation sigma lambda}
\sum_{\lambda'}\Tr [\sigma_{\lambda'}]&= \sum_{\lambda'}\int d\lambda \mu(\lambda) p(\lambda'|\lambda),\nonumber\\
&= 1.
\end{align}
}

We can now easily write a SDP that tests if a given assemblage $\sigma_{a|x}$ with $x=1,...,m_\rA$ and $a=1,...,o_\rA$ is LHS \cite{P13}:
\begin{align}
\text{given}& \quad\{\sigma_{a|x}\}_{a,x},\{D(a|x,\lambda)\}_{\lambda} \nonumber \\
\text{find}& \quad \left\{ \sigma_\lambda\right\}_{\lambda} \label{e:feasibility} \\
\text{s.t.}& \quad \sum_\lambda D(a|x,\lambda)\sigma_\lambda = \sigma_{a|x}  \quad\forall a,x,\nonumber \\
&\quad \sigma_\lambda \geq 0\quad \forall \lambda. \nonumber
\end{align}
Notice first that the constraint \eqref{e:normalisation sigma lambda} is implicitly enforced in the above, due to condition \eqref{normalization} satisfied by all quantum assemblages.  Furthermore, notice that this problem is a special instance of a generic SDP where the objective function \emph{vanishes} (i.e. $A_\lambda = 0$, in the terminology of Appendix~\ref{SDP}, such that the objective is $\sum_\lambda \tr[A_\lambda \sigma_\lambda] = 0$). Such problems are called \emph{feasibility problems}, and test whether or not the (primal) feasible set is equal to the empty set or not; if it is equal to the empty set there exists no set of $\sigma_\lambda$ which satisfy the constraints in \eqref{e:feasibility}. In the cases where the set is not empty, the optimal primal value is zero. If on the other hand the set is empty, the primal optimal value is $\alpha = -\infty$, i.e. as small as possible, which indicates the infeasibility of the problem.

Finally, let us also note that we can always turn this feasibility SDP into a strictly feasible one, which is advantageous from a computational perspective. In particular, if we relax the constraint $\sigma_\lambda \geq 0 \qu\forall \lambda$ to $\sigma_\lambda \geq \mu \openone\qu \forall \lambda$, (where $\mu$ can be negative), then there is always a $\mu$ such that $\sum_\lambda D(a|x,\lambda) \sigma_\lambda = \sigma_{a|x}\qu \forall a,x$\footnote{This follows from the fact that any assemblage can be bought inside the set of LHS assemblages by mixing with the maximally mixed assemblage $\openone_{a|x} := \openone/(d_\rB o_\rA)$ (\ie there is a ball of LHS assemblages around it). Re-arranging such a mixture demonstrates the claim.}. Thus, one can equivalently test for an LHS model by solving the following SDP:
\begin{align}
\text{given}& \quad\{\sigma_{a|x}\}_{a,x},\{D(a|x,\lambda)\}_{\lambda} \nonumber \\
\max_{\{\sigma_\lambda\}}&\quad \mu \label{e:LHS feas} \\
\text{s.t.}& \quad \sum_\lambda D(a|x,\lambda)\sigma_\lambda = \sigma_{a|x}  \quad\forall a,x,\nonumber \\
&\quad \sigma_\lambda \geq \mu \openone\quad \forall \lambda. \nonumber
\end{align}
A negative optimal value demonstrates steering, as this corresponds to the cases when \eqref{e:feasibility} was infeasible. On the other hand, a non-negative value means that all $\sigma_\lambda$ are positive semi-definite (PSD), and thus corresponds to having obtained an LHS model (\ie to the feasible case above).

\subsection{Optimal steering inequalities}\label{Sec: dual}
We can now apply the duality theory of SDPs to the above problem \eqref{e:LHS feas} to obtain the dual problem. Following the steps outlined in Appendix~\ref{SDP}, that is by introducing dual (Lagrange) variables, and passing to the Lagrangian, we arrive at the  following dual program:
\begin{align}\label{e:dual}
\text{given} &\quad \{\sigma_{a|x}\}_{a,x}, \{D(a|x,\lambda)\}_\lambda \nonumber \\
\min_{\{F_{a|x}\}} & \quad\tr\sum_{ax}F_{a|x}\sigma_{a|x} \\
\text{s.t.} &\quad \sum_{ax}F_{a|x}D(a|x,\lambda) \geq 0 \quad \forall \lambda,\nonumber \\
&\quad \tr\sum_{ax,\lambda} F_{a|x} D(a|x,\lambda) = 1. \nonumber
\end{align}

If $\sigma_{a|x}$ demonstrates steering, the solution of this problem returns Hermitian operators $\{F_{a|x}\}_{ax}$, which can be used to define a steering inequality $\quad\tr\sum_{ax}F_{a|x}\sigma_{a|x}\geq\beta_{\LHS}$ which is satisfied by all LHS assemblages and is violated, in particular, by $\sigma_{a|x}$. The first constraint encodes the requirement that all LHS assemblages should achieve a value larger than 0: indeed, by multiplying both sides by the members of an LHS model, $\sigma_\lambda$, and taking the sum and trace, we see that $\tr \sum_{ax\lambda}F_{a|x}D(a|x,\lambda) \sigma_\lambda \geq 0$. That is, $\beta_\LHS=0$. On the other hand, the dual objective function is the value $\beta := \tr\sum_{ax}F_{a|x}\sigma_{a|x}$ obtained by the observed assemblage. Only if the assemblage demonstrates steering can this value be negative. Finally, the second constraint can be seen as fixing a scale for the steering functional, in particular by specifying the value taken for the LHS model $\sigma_\lambda = \openone/(d_\rB|\lambda|)$.

\subsubsection{Example: the 2-qubit Werner state}
As an illustration of the above, we demonstrate here how the optimal steering inequality can be obtained from the dual \eqref{e:dual} in one of the simplest steering scenarios: Alice performs two Pauli spin measurements ($X$ and $Z$) on half of the two-qubit Werner state,
\be\label{e: werner state}
\rho(w)=w\ket{\phi^+}\bra{\phi^+}+(1-w)\frac{\openone}{4},
\ee
where $\ket{\phi^+}=(\ket{00}+\ket{11})/\sqrt{2}$ and $0\leq w \leq1$. {It is well known that this state is steerable if and only if $w>1/2$  \cite{WJD07}. However, to demonstrate steering up to this value of $w$ an infinite number of projective measurements have to be used.} For the simple case at hand, it is known that the state is steerable for two measurements if $w > 1/\sqrt{2}$ \cite{CJWR09}.

The assemblage created for Bob is given by
\begin{align}\label{e:Werner assem}
\sigma_{a|x} = w\frac{\openone + (-1)^a \hat{n}_x\cdot \vec{\sigma}}{2} + \frac{1-w}{4}\openone
\end{align}
where $\vec{\sigma}=(X,Y,Z)$ is a vector containing the Pauli operators, and $\hat{n}_0 =\hat{x}$ and $\hat{n}_1 = \hat{z}$ are unit vectors on the Bloch sphere.

An arbitrary Hermitian $F_{a|x}$ can be written as {$F_{a|x} = \alpha_{a|x} \openone+ \vec{m}_{a|x}\cdot \vec{\sigma}$}, for $\alpha_{a|x}$ a real scalar, and $\vec{m}_{a|x}$ a real three-dimensional vector. For this steering functional, the assemblage \eqref{e:Werner assem} achieves the value
\begin{multline}\label{e:werner beta}
\beta = \frac{1}{2}(\alpha_{0|0} + \alpha_{1|0} + \alpha_{0|1} + \alpha_{1|1}) \\
+ \frac{w}{2}(\hat{n}_0\cdot (\vec{m}_{0|0} - \vec{m}_{1|0}) + \hat{n}_1\cdot(\vec{m}_{0|1} - \vec{m}_{1|1})).
\end{multline}

{Recall that $\lambda$ runs over all deterministic strategies, $\lambda = (a_{x=0},a_{x=1})$}. The first constraint in \eqref{e:dual} when $\lambda = (0,0)$, namely $F_{0|0} + F_{0|1} \geq 0$, written out explicitly is
\begin{equation}
(\alpha_{0|0} + \alpha_{0|1})\openone + (\vec{m}_{0|0} + \vec{m}_{0|1})\cdot \vec{\sigma} \geq 0,
\end{equation}
which is easily seen to be equivalent to the condition $(\alpha_{0|0} + \alpha_{0|1}) \geq \| \vec{m}_{0|0} + \vec{m}_{0|1}\|$. The same calculation for the other three cases in \eqref{e:dual} (when $\lambda = (0,1)$, $(1,0)$ and $(1,1)$ respectively) lead to the four constraints:
\begin{equation}
\begin{split} \label{e:werner ex constr}
(\alpha_{0|0} + \alpha_{0|1}) \geq \| \vec{m}_{0|0} + \vec{m}_{0|1}\|,  \\
(\alpha_{0|0} + \alpha_{1|1}) \geq \| \vec{m}_{0|0} + \vec{m}_{1|1}\|, \\
(\alpha_{1|0} + \alpha_{0|1}) \geq \| \vec{m}_{1|0} + \vec{m}_{0|1}\|, \\
(\alpha_{1|0} + \alpha_{1|1}) \geq \| \vec{m}_{1|0} + \vec{m}_{1|1}\|.
\end{split}
\end{equation}
The second constraint is also straightforwardly seen to be equivalent to $(\alpha_{0|0} + \alpha_{1|0} + \alpha_{0|1} + \alpha_{1|1}) = 1/4$.

Now, it is clear that the second line of \eqref{e:werner beta} is minimised when we anti-align (or align) the $\vec{m}_{a|x}$ with the $\hat{n}_x$, \ie when $\vec{m}_{0|0} = -m_{0|0} \hat{n}_0$, $\vec{m}_{1|0} = m_{1|0} \hat{n}_0$, $\vec{m}_{0|1} = -m_{0|1} \hat{n}_1$, and $\vec{m}_{1|1} = m_{1|1} \hat{n}_1$, for $m_{a|x}$ positive scalars. Now, let us consider the simple symmetric case as an ansatz: $\alpha_{a|x} = \alpha = 1/16$, $m_{a|x} = m$. In this case, the constraints \eqref{e:werner ex constr} become identical, and are saturated when $m = 1/(8\sqrt{2})$. Putting everything together, we see that $\beta = 1 - \sqrt{2}w$, and thus we have a violation whenever $w > 1/\sqrt{2}$.

Finally, defining $A_x := M_{0|x} - M_{1|x}$, the (unknown) observable measured by Alice, such that $\tr_\rA[(A_x \otimes \openone)\rho^{\rA\rB}] = \sigma_{0|x} - \sigma_{1|x}$, it is straightforward to show that
\begin{equation}
\beta = \tr\sum_{ax}F_{ax}\sigma_{a|x} = \tfrac{1}{16}(2-\sqrt{2}\langle A_0 X \rangle  -\sqrt{2}\langle A_1 Z \rangle )
\end{equation}
where $\langle A_0 F \rangle := \tr[(\sigma_{0|x} - \sigma_{1|x})F]$, which leads to the steering inequality
\begin{equation}
\langle A_0 X \rangle + \langle A_1 Z \rangle \leq \sqrt{2}.
\end{equation}
This coincides with the well-known linear steering inequality first derived in \cite{CJWR09}. A similar analysis also works for the case where Alice performs all 3 Pauli spin measurements $X$, $Y$, and $Z$ on the Werner state, in which case the optimal linear steering inequality $\langle A_0 X \rangle + \langle A_1 Y \rangle + \langle A_2  Z \rangle \leq \sqrt{3}$ is obtained.

\subsection{Further results on steering detection}

We have described above how to detect whether an assemblage demonstrates steering by semidefinite programming. There are several  other results on steering detection, most of them based on developing specific steering inequalities. Here we review some of these studies.

To the best of our knowledge, the first proposal of a steering inequality was made in the continuous-variable case \cite{R89}. We leave the discussion of this scenario to the Sec.~\ref{cv inequalities}.

Several linear steering inequalities were proposed in Refs. \cite{CJWR09,SJWP10,BES+12,ECW13,EW14,SC15}. Some of these inequalities were proven to be useful in experimental steering tests with inefficient detectors \cite{BES+12,ECW13,EW14,SC15}. In particular, Ref.~\cite{BES+12} proposed a steering inequality which is violated if Alice has up to 16 qubit measurements with detection efficiency above $1/m_\rA$, where $m_\rA$ is the number of measurements (see also \cite{ECW13,EW14}). This result was generalised for any number of arbitrary measurements, of any dimension, in Ref.~\cite{SC15}. {We notice that this detection efficiency threshold is tight, since if the detection efficiency of Alice's measurements is below this threshold no steering can be demonstrated ~\cite{BES+12}.}

Another interesting class of linear steering inequalities are those allowing for an unbounded violation \cite{HMY14,MRY+15,SC15}. In Ref.~\cite{MRY+15} the authors showed a steering inequality whose violation increases indefinitely with the dimension of the system measured if Alice performs measurements given by mutually unbiased bases. In Ref.~\cite{SC15} examples of unbounded violations in a more realistic scenario consisting of more general measurements and errors in the state preparation and measurements were shown (see also \cite{RBHS16}).

Concerning the relation between steering and joint measurability, Refs.~\cite{QVB14,UMG14} showed that any set of incompatible measurements lead to steering when applied to any entangled pure state.

Finally, modifications to the steering scenario were also considered. In Refs.~\cite{QZFY16,CFFW15,RBMB15,GC16} a scenario where Alice and Bob are restricted to perform only two dichotomic measurements each was considered. Ref.~\cite{MGHG14} considered the case where Bob, instead of performing tomography, only assumes that his subsystem has a given dimension. In Ref.~\cite{CHW13}  the tomography in Bob's site is changed by uncharacterised measurements with known quantum inputs. {In Ref.~\cite{CYW+13} a method to detect steering without the use of inequalities was proposed. Ref.~\cite{CSS16} considered that Alice has trusted quantum inputs, while Bob applies tomography, and showed the connection of this scenario with that of quantum teleportation.}

\section{Quantum states with local-hidden-state models}
\label{Sec:LHS states}

In Sec.~\ref{Sec: detecting} we have described ways of checking if a given assemblage demonstrates steering. However there exists entangled states that are unsteerable, \ie all assemblages obtained by measuring them are LHS \cite{WJD07,APB+07,BVQB14,QVC+15}. An example of such a state is the two-qubit Werner state \eqref{e: werner state} for which if Alice applies any projective measurement the resulting assemblage is LHS if $p\leq1/2$ (this follows from the fact that the model presented in Ref.~\cite{W89} is a LHS model \cite{WJD07,APB+07}).  Notice that above this bound this state is known to be steerable \cite{WJD07}. For general POVM measurements, this state produces LHS assemblages if $p\leq5/12$ (this follows from the fact that the local hidden variable (LHV) model presented in Ref.~\cite{B12} can again also be seen as a LHS model \cite{QVC+15}). This bound is however not known to be tight. 

Determining whether a given state is unsteerable or not is an important issue since it tells us whether the entanglement of such a state can be certified if one of the parties uses uncharacterised devices. Moreover, it also identifies whether the state is useful for steering based tasks (see Sec.~\ref{Sec: applications}). Additionally, unsteerable states can not violate Bell inequalities. This follows from the fact that the probability distributions that are obtained after Bob applies a local measurement on a LHS assemblage \eqref{LHS} have a LHV model \cite{BCP+14}, \ie:
\begin{align}
p(ab|xy)&=\tr[M_{b|y}\sigma_{a|x}^{\LHS}],\nonumber \\
&=\tr[M_{b|y}\int d\lambda \mu(\lambda) p(a|x,\lambda) \rho_\lambda],\nonumber\\
&=\int d\lambda \mu(\lambda) p(a|x,\lambda)p(b|y,\lambda).
\end{align}
Thus, unsteerable states are also local, and hence useless for device-independent tasks \cite{BCP+14}.

Unfortunately determining if a state is unsteerable is a very hard problem, because it involves determining if the assemblages generated from it have the LHS form \eqref{LHS} for all possible measurements Alice can apply. However recent techniques based on SDP have provided a powerful sufficient test to determine whether quantum states are unsteerable \cite{CGRS15,HQV+15}.  There are two variants of this test. In the first, given a specific state, a test was given which can certify its unsteerability \cite{CGRS15,HQV+15}. The second variants allows one to probe whether there exists an unsteerable state in a given region of the state space \cite{CGRS15}. Thus by probing different regions one can find different unsteerable states. In what follows we describe these methods.

\subsection{From infinite to finite measurements}\label{sec:infinite to finite}

The main difficulty in determining if a quantum state is unsteerable relies in the fact that infinitely many measurements have to be considered. In what follows we outline a method which gets rid of this problem by replacing it with one of finding a LHS model for a finite number of measurements on a (potentially unphysical) state. {This method was inspired by the results of Ref.~\cite{BHQB15} where the authors propose LHV/LHS models using only a finite number of hidden variables/states}, and will allow the use of semidefinite programming in situations which involve infinitely many measurements, which \textit{a priori} would appear to be beyond the scope of SDP techniques. Here we will restrict our discussion to qubits and projective measurements, although a similar discussion could be made for more general cases. 

Consider a set of $m_\rA$ projective measurements $\{M_x\}_{x=0}^{m_{\rA}-1}$, each of them with measurement operators $\Pi_{a|x}=[\openone+(-1)^a\vec{u}_{x} \cdot \vec{\sigma}]/{2}$, with $a\in \{0,1\}$, $\vec{u}_x$ unit vectors, and $\vec{\sigma}=(X,Y,Z)$ a vector containing the Pauli operators. These measurements can be represented by points on the surface of the Bloch sphere, and define a polytope in $\mathbb{R}^3$. We can calculate the radius $r$ of the largest ball contained in this polytope\footnote{This can be done by enumerating the facets of the polytope, and looking for the one that is closet to the center of the ball.}
 which will represent noisy measurements with elements $\Pi_{a|\vec{u}}^{(r)}=r\Pi_{a|\vec{u}}+(1-r)\openone/2$, where $\vec{u}$ are arbitrary unit vectors in $\mathbb{S}^3$. This implies that the entire (infinite) ball of noisy measurements can be simulated using only the finite set of measurements $\{M_x\}_x$.

Now, suppose that the set of measurements $\{M_x\}_x$, when applied to a Hermitian operator $O^{\rA\rB}$, leads to a LHS assemblage, \ie:
\begin{equation}
\tr_\rA[(\Pi_{a|x}\otimes\openone) O^{\rA\rB}]=\sum_\lambda D(a|x,\lambda)\sigma_\lambda\quad\forall~a,x.
\end{equation}
Then, the identity
\begin{multline}
\tr_\rA[(\Pi_{a|\hat{u}}^{(r)}\otimes \openone) O^{\rA\rB} ]\\= \tr_\rA[( \Pi_{a|\hat{u}}\otimes \openone) (rO^{\rA\rB}+(1-r)\frac{\openone}{2}\otimes O^{\rB})],
\end{multline}
where $O^{\rB}=\tr_\rA[ O^{\rA\rB}]$, implies that the operator $O^{\rA\rB'}=r O^{\rA\rB}+(1-r)\openone/2\otimes O^{\rB}$ has a LHS model for all projective measurements $\Pi_{a|\hat{u}}$.

More generally,  if one can find a set of measurements $\{M_x\}_x$, and a linear map $\Lambda(\cdot)$ which shrinks the Bloch sphere so that it is contained within the convex hull of $\{M_x\}_x$, then the identity 
\begin{equation}\tr_\rA[(\Lambda(\Pi_{a|\hat{u}})\otimes \openone) O^{\rA\rB} ]\\= \tr_\rA[( \Pi_{a|\hat{u}}\otimes \openone) \Lambda^\dagger\otimes\mathrm{id}(O^{\rA\rB})],
\end{equation}
where $\mathrm{id}(\cdot)$ refers to the identity channel, implies that $\Lambda^\dagger\otimes\mathrm{id}(O^{\rA\rB})$ will have a LHS model for all projective measurements whenever $O^{\rA\rB}$ does for the measurements $\{M_x\}_x$.

\subsection{Target states}
We can now present a method to test if a given state $\rho^{\rA\rB}$ in $\mathbb{C}^2\otimes\mathbb{C}^d$ is unsteerable when Alice applies projective measurements \cite{CGRS15,HQV+15}. The key ingredient is the method outlined in the previous section, which allows one to determine whether a noisy state has a LHS model for all projective measurements, by checking whether the operator $O^{\rA\rB}$ has a LHS model for a finite set of fixed measurements $\{M_x\}_{x=0}^{m_\rA-1}$ with measurement operators $\{\Pi_{a|x}\}_{a=0}^{a=o_\rA-1}$. 

With this in mind we can now define a SDP which, for any choice of measurements $M_x$, searches for an operator $O^{\rA\rB}$ which provides a LHS assemblage for these measurements, and such that $rO^{\rA\rB}+(1-r)\openone/2\otimes O^{\rB}=\rho^{\rA\rB}$
\begin{align}
\text{given}&\quad\rho^{\rA\rB},\{\Pi_{a|x}\}_{a,x},r \nonumber \\
\text{find} &\quad O^{\rA\rB},\{\sigma_\lambda\}_\lambda   \label{eq:lhs-proj-fixed-state} \\
\text{s.t.}&\quad\tr_\rA\left[\left(\Pi_{a|x}\otimes \openone \right)O^{\rA\rB}\right]=\sum_\lambda D(a|x,\lambda)\sigma_\lambda \quad\forall~a,x,  \nonumber  \\
 &\quad \sigma_\lambda\geq0 \quad\forall~\lambda, \nonumber\\
 &\quad rO^{\rA\rB}+(1-r)\openone/2\otimes O^{\rB}=\rho^{\rA\rB}. \nonumber
\end{align}
The fact that the resulting $O^{\rA\rB}$ is unsteerable for the set $\{M_x\}_x$ implies that $\rho^{\rA\rB}$ is unsteerable for \textit{all} projective measurements. Notice that the operator $O^{\rA\rB}$ is not required to be a physical state. The only requirement is that it becomes the desired state $\rho^{\rA\rB}$ after white noise is applied on half of it (a similar idea was used in Ref.~\cite{AJRV14}).

In Refs.~\cite{CGRS15,HQV+15} it was shown that this method can detect that several entangled quantum states are unsteerable including Werner states \eqref{e: werner state} and some variations of it, $2 \times 4$ bound entangled states, noisy GHZ and W states \eqref{eq: noisy tripartite}.

In order to also include the case where Alice applies general POVM measurements, use of a result in Ref.~\cite{HQB+13} can be made, proving that if a state $\rho^{\rA\rB}$  is unsteerable for all projective measurements, then the state $\rho^{\rA\rB'}=(1/2)\rho^{\rA\rB}+(1/2)\gamma^\rA\otimes\rho^\rB$ has a LHS model for general POVM measurements, where $\gamma^\rA$ is an arbitrary state. The above SDP can be modidied to include this additional noise:
\begin{align}
\text{given} &\quad\rho^{\rA\rB},\{\Pi_{a|x}\}_{a,x},r, \gamma^\rA \nonumber \\
\text{find} &\quad O^{\rA\rB},\{\sigma_\lambda\}_\lambda   \label{eq:lhs-povm-fixed-state} \\
\text{s.t.} &\quad \tr_\rA\left[\left(\Pi_{a|x}\otimes \openone \right)O^{\rA\rB}\right]=\sum_\lambda D(a|x,\lambda)\sigma_\lambda\quad \forall~a,x,  \nonumber \\
 &\quad \sigma_\lambda\geq0 \quad\forall~\lambda,\nonumber\\
 &\quad \frac{1}{2}\left[ r O^{\rA\rB}+(1-r)\frac{\openone}{2}\otimes O^{\rB} \right]+\frac{\gamma^\rA \otimes O^\rB}{2}=\rho^{\rA\rB}. \nonumber
\end{align}

\subsection{Witness-generated states}

In the previous SDPs  \eqref{eq:lhs-proj-fixed-state} and \eqref{eq:lhs-povm-fixed-state} the starting point is a target state $\rho^{\rA\rB}$. In Ref.~\cite{CGRS15} a variant of this method was proposed to randomly find  entangled states which are unsteerable. The idea is to start with an entanglement witness $W$ (\ie an operator for which $\tr[W\rho^{\rA\rB}]<0$ implies that $\rho^{\rA\rB}$ is entangled) and look for unsteerable states that can be detected by it. The new SDPs read
\begin{align}
\text{given}&\quad W,\{\Pi_{a|x}\}_{a,x},r \nonumber \\
 \min_{O^{\rA\rB}, \{\sigma_\lambda\}}& \quad\tr[W \tilde{O}^{\rA\rB}] \label{eq:sdp-proj-random-state}  \\
 \text{s.t.} &\quad \tilde{O}^{\rA\rB}=rO^{\rA\rB}+(1-r)\openone /2\otimes O^{\rB},\nonumber\\
&\quad \tilde{O}^{\rA\rB}\geq 0, \quad \tr [\tilde{O}^{\rA\rB}] = 1, \nonumber \\
  &\quad\tr_\rA\left[\left(\Pi_{a|x}\otimes \openone \right)O^{\rA\rB}\right]=\sum_\lambda D(a|x,\lambda)\sigma_\lambda\qu \forall~a,x, \nonumber  \\
 &\quad\sigma_\lambda\geq0 \quad\forall~\lambda,\nonumber
\end{align}
in the case of LHS for projective measurements and
\begin{align}
\text{given}&\quad W,\{\Pi_{a|x}\}_{a,x},r,\gamma^\rA \nonumber  \\
 \min_{O^{\rA\rB},\{\sigma_\lambda\}}&\quad\tr[W\tilde{O}^{\rA\rB}] \label{eq:sdp-povm-random-state} \\
 \text{s.t.} &\quad \tilde{O}^{\rA\rB}=\tfrac{1}{2}(rO^{\rA\rB}+(1-r)\tfrac{\openone}{2}\otimes O^{\rB})+\tfrac{1}{2}(\gamma^\rA\otimes O^\rB),\nonumber\\
 &\quad \tilde{O}^{\rA\rB}\geq 0, \quad \tr [\tilde{O}^{\rA\rB}] = 1, \nonumber\\
 &\tr_\rA\left[\left(\Pi_{a|x}\otimes \openone \right)O^{\rA\rB}\right]=\sum_\lambda D(a|x,\lambda)\sigma_\lambda\quad \forall~a,x,   \nonumber\\
 &\sigma_\lambda\geq 0 \quad\forall~\lambda,\nonumber
\end{align}
for the case of POVM measurements. If the solution of these SDPs are negative, then the minimising operator $\tilde{O}^{\rA\rB}$ is an entangled state which is unsteerable: The fact that $\tilde{O}^{\rA\rB}$ is entangled is guaranteed by the violation of the witness $W$ and the facts that it is a valid state and has a LHS model are imposed by the constraints of the SDPs.

Using the above SDPs for randomly chosen entanglement witnesses, thousands  of new unsteerable entangled states were found \cite{CGRS15}. 

\subsection{Further results on LHS models}

{Even before quantum steering gained attention, there were some entangled quantum states which were known to not be steerable. This is because some LHV models can be also seen as LHS models. This is the case, as mentioned before, of the celebrated LHV model for projective measurements applied on the Werner state shown in Ref. \cite{W89}. Another example is the LHV model for general POVM measurements for the same state presented by Barrett \cite{B12} (see \cite{QVC+15}). Similar ideas were used to present LHS models for other classes of states in Refs.~\cite{WJD07,APB+07,HQB+13}. Furthermore, in the multipartite case, there are also known examples of (genuine) multipartite entangled states with LHS models \cite{TA06,AAD+15,BFF+16}.}

Concerning other criteria to certify whether quantum states are unsteerable we suggest Refs.~\cite{JHA+15,BHQB15}. In particular Ref.~\cite{JHA+15}, following the techniques of Ref.~\cite{JPJR14}, described a sufficient condition for a Bell-diagonal state to be unsteerable when projective measurements are considered, which was later proven to be also necessary \cite{CV16}. In Ref.~\cite{BHQB15a} a different sufficient criterion, that can be applied to any two-qubit state, was found.

\section{Quantification of quantum steering}
\label{Sec: quantification}

In the Section~\ref{Sec: detecting} we reviewed the problem of detecting quantum steering from a given assemblage. Another interesting question recently considered in the literature is how much steering does an assemblage demonstrate. In what follows we will review some of the recently proposed quantifiers of steering which can be calculated using semidefinite programming. These quantifiers were motivated by entanglement and nonlocality quantifiers and have clear operational interpretations.

\subsection{The steering weight}

The steering weight was the first proposed quantifier of steering \cite{SNC14} and was motivated by the best separable approximation of entanglement \cite{LS98} and the EPR2 decomposition of nonlocality \cite{EPR92}. Consider an assemblage $\sigma_{a|x}$. We can always decompose it as a convex combination of a LHS assemblage $\sigma_{a|x}^{\LHS}$ of the form \eqref{LHS} and a generic assemblage $\gamma_{a|x}$ satisfying \eqref{nosignaling} and $\eqref{normalization}$, \ie
\begin{equation}\label{eq: SW decomposition}
\sigma_{a|x}=p\gamma_{a|x}+(1-p)\sigma_{a|x}^{\LHS}\quad\forall a, x.
\end{equation}
This decomposition can be interpreted as if the assemblage $\sigma_{a|x}$ is being created by mixing members of the assemblage $\gamma_{a|x}$ and a LHS assemblage $\sigma_{a|x}^{\LHS}$, with weights $p$ and $1-p$ respectively. The steering weight of an assemblage $\sigma_{a|x}$, denoted by $\SW(\sigma_{a|x})$, quantifies how much of the assemblage $\gamma_{a|x}$ is needed in such a procedure (consequently, how much of a LHS assemblage can be used). That is, it consists in the minimum of $p$ over all possible decompositions of the form \eqref{eq: SW decomposition}:
\begin{align}\label{eq: SW}
\SW(\sigma_{a|x})= \!\!\!\!\min_{p,\{\gamma_{a|x}\},\{\sigma_\lambda\}}\!\!\!\! &\quad p\\
\text{s.t.} &\quad \sigma_{a|x}=p\gamma_{a|x}+(1-p)\sigma_{a|x}^{\LHS}\quad\forall a, x,\nonumber\\
&\quad\sigma_{a|x}^{\LHS}=\sum_\lambda D(a|x,\lambda)\sigma_\lambda\quad\forall a,x,\nonumber\\
&\quad\sum_a \gamma_{a|x}=\sum_a \gamma_{a|x'}\quad\forall x,x',\nonumber\\
&\quad\tr\sum_a \gamma_{a|x}=1\quad\forall x,\quad\gamma_{a|x}\geq 0\quad\forall a,x,\nonumber \\
&\quad\tr\sum_\lambda \sigma_\lambda=1,\quad\sigma_\lambda\geq 0\quad\forall \lambda,\nonumber
\end{align}
where we have used Eq.~\eqref{deterministicLHS} to write $\sigma^\LHS_{a|x}$ as a combination of deterministic strategies.

An interesting fact about the steering weight is that it is bounded by any convex function $f(\cdot)$ of the assemblage. For instance, $f(\cdot)$ can be the violation of a steering inequality. Let us consider that the steering weight of the assemblage $\sigma_{a|x}$, given by $p^*=\SW(\sigma_{a|x})$, is achieved by the decomposition $\sigma_{a|x}=p^*\gamma^*_{a|x}+(1-p^*)\sigma_{a|x}^{*\LHS}$. Then,
\begin{align}
f(\sigma_{a|x})&\leq p^*f(\gamma^*_{a|x})+(1-p^*)f(\sigma_{a|x}^{*\LHS}),\nonumber\\
&\leq p^* f_{\max}+(1-p^*)f_{\max}^{\LHS},
\end{align}
where in the first line we used the convexity of $f(\cdot)$, and in the second line we denote the maximal value of $f(\cdot)$ among all possible assemblages by $f_{\max}$, and among all possible LHS assemblages by $f_{\max}^{\LHS}$. This implies that
\begin{align}
\SW(\sigma_{a|x})\geq \frac{f(\sigma_{a|x})-f_{\max}^{\LHS}}{f_{\max}-f_{\max}^{\LHS}}.
\end{align}

It is not immediately evident from the definition \eqref{eq: SW} that the computation of SW can be made via SDP since $p$, $\sigma_{a|x}^{\LHS}$ and $\gamma_{a|x}$ are all variable of the problem. However, after some manipulation we can rewrite this problem as such. In order to see this, let us first combine the condition that $\gamma_{a|x} \geq 0$ with the decomposition for $\sigma_{a|x}$ to obtain
\begin{equation}
	\gamma_{a|x} = \tfrac{1}{p}\left(\sigma_{a|x} - (1-p) \sum_\lambda D(a|x,\lambda)\sigma_\lambda \right) \geq 0.
\end{equation}
Without loss of generality we can take $p > 0$ ($p=0$ means that the assemblage is LHS), then the term inside the brackets must be positive semidefinite. Notice that the condition $\sum_a \gamma_{a|x} = \sum_a \gamma_{a|x'}$ does not need to be explicitly enforced, since by assumption $\sigma_{a|x}$  satisfies
\begin{equation}
\begin{aligned}
	\sum_a \sigma_{a|x} &= \sum_a \sigma_{a|x'} \quad\quad \forall x, x'.
\end{aligned}
\end{equation}

Defining new variables $\tilde{\sigma}_\lambda = (1-p) \sigma_\lambda$, we have that
\begin{align}
\tr \sum_\lambda \tilde{\sigma}_\lambda = 1-p,
\end{align}
since $\sum_a D_\lambda(a|x) = 1$ for all $x$ (as they are valid probability distributions).
Combining this with the above, we finally arrive at the final form for the SDP:
\begin{align} \label{e:weight SDP simple}
\SW(\sigma_{a|x}) = \min_{\{\tilde{\sigma}_\lambda\}}& \quad 1-\Tr \sum_\lambda\tilde{\sigma}_\lambda \\
\text{s.t.}& \quad \sigma_{a|x} - \sum_\lambda D(a|x,\lambda)\tilde{\sigma}_\lambda \geq 0 \quad \forall a,x, \nonumber \\
	& \quad \tilde{\sigma}_\lambda \geq 0 \quad \forall \lambda. \nonumber
\end{align}

Using the duality theory presented in Appendix \ref{SDP}, one can also express the steering weight as a maximisation over steering inequalities of a certain type \cite{SNC14}. In particular, the dual formulation of \eqref{e:weight SDP simple} is
\begin{equation}\label{e:weight SDP dual final}
\begin{aligned}
\SW(\sigma_{a|x}) = \max_{\{F_{a|x}\}}& \quad 1-\Tr \sum_{ax} F_{a|x}\sigma_{a|x} \\
\text{s.t.}& \quad 0 \geq \openone - \sum_{ax} D(a|x,\lambda)F_{a|x} \quad \forall \lambda, \\
	& \quad F_{a|x} \geq 0 \quad \forall a,x.
\end{aligned}
\end{equation}
Crucially, strict duality is seen to hold, since the dual is strictly feasible: $F_{a|x} = \alpha \openone \qu \forall a,x$, for $\alpha > 0$ strictly satisfies all the constraints for a sufficiently large $\alpha$.

In order to interpret this SDP let us notice that any LHS assemblage $\sigma^{\LHS}_{a|x}=\sum_\lambda D(a|x,\lambda)\sigma_\lambda$ ($\Tr\sum_\lambda \sigma_\lambda = 1$ and $\sigma_\lambda \geq 0$) satisfies
\begin{equation}
1 - \Tr\sum_{ax} \sigma_{a|x}^\LHS F_{a|x}\leq 0.
\end{equation}
This can be seen by multiplying  $\openone - \sum_{ax} D_\lambda(a|x)F_{a|x}$ by $\sigma_\lambda$ and taking the sum and trace. Thus $\Tr\sum_{ax}F_{a|x}\sigma_{a|x}^\LHS\geq 1$. On the other hand, since $F_{a|x}\geq0$ we have that $\beta = \tr\sum_{ax}F_{a|x} \sigma_{a|x} \geq 0$. In total, this defines a positive linear steering functional with local bound $\beta_\LHS = 1$, and the program \eqref{e:weight SDP dual final} searches for the optimal such inequality for the provided assemblage. 

\subsection{Robustness-based steering quantifiers}

A similar approach for the quantification of steering is to ask how much noise one has to add to a given assemblage in order for it to have an LHS model, analogously to the robustness of entanglement \cite{VT99}. In general terms the $\mathcal{N}$-Robustness of an assemblage can defined as
\begin{align} \label{eq: robustness}
\SR^{\mathcal{N}}(\sigma_{a|x}) = \min_{\{\pi_{a|x}\},\{\sigma_\lambda\},t} & \quad \!t \\
\text{s.t.}& \quad \frac{\sigma_{a|x} + t \pi_{a|x}}{1+t} = \sigma_{a|x}^\LHS \quad\forall a,x, \nonumber\\
	& \quad \sigma_{a|x}^\LHS = \sum_\lambda D(a|x,\lambda)\sigma_\lambda \quad\forall a,x, \nonumber \\
	& \quad \pi_{a|x} \in \mathcal{N}, \quad \sigma_\lambda \geq 0 \quad\forall \lambda,\nonumber
\end{align}
where $\mathcal{N}$ is any (convex) subset of assemblages characterised by {positive semi-definite (PSD) constraints and linear matrix inequalities (LMIs)}. This set will determine the specific type of noise one is interested in, and the corresponding robustness quantifier. Notice that we have chosen to express the convex combination above in terms of the weights $[1/(t+1),t/(t+1)]$, instead of the usual $[p,1-p]$. This is simply to make the robustness of steering analogous to the robustness of entanglement proposed in \cite{VT99}.

As with the steering weight, such robustness quantifiers are not explicitly in the form of a semidefinite program, however they can always be re-expressed explicitly as one. In particular, first note that we can always take $t > 0$, since $t = 0$ corresponds to $\sigma_{a|x}$ being LHS. Then, combining the first and second constraints with $\pi_{a|x} \in \mathcal{N}$, we obtain the equivalent condition
\begin{equation}
\pi_{a|x} = \frac{1}{t}\left((1+t) \sum_\lambda D(a|x,\lambda) \sigma_\lambda - \sigma_{a|x}\right) \in \mathcal{N}.
\end{equation}
Now, defining new variables $\tilde{\sigma}_{\lambda} = (1+t)\sigma_\lambda$  we have that $\tr\sum_\lambda \tilde{\sigma}_\lambda = (1+t)$. Finally, we define a new set $\tilde{\mathcal{N}} = \{\tilde{\pi}_{a|x} | \tilde{\pi}_{a|x} = t\pi_{a|x}, \pi_{a|x} \in \mathcal{N}, t\geq 0\}$. This set is simply the conic hull of $\mathcal{N}$. Finally, note that $\tilde{\mathcal{N}}$ maintains a characterisation in terms of PSD constraints and LMIs\footnote{This follows because (i) multiplying by a positive scalar doesn't change the positivity of operators (ii) linear matrix inequalities remain linear matrix inequalities after multiplying by a scalar.}.

We thus arrive at the SDP formulation of \eqref{eq: robustness}, given by
\begin{align} \label{eq: robustness final}
\SR^{\mathcal{N}}(\sigma_{a|x}) = \min_{\{\tilde{\sigma}_\lambda\}} & \quad \tr\sum_\lambda \tilde{\sigma}_\lambda-1 \\
\text{s.t.}& \quad  \sum_\lambda D(a|x,\lambda)\tilde{\sigma}_\lambda - \sigma_{a|x} \in \tilde{\mathcal{N}}, \nonumber\\
	& \quad \tilde{\sigma}_\lambda \geq 0 \quad\forall \lambda.\nonumber
\end{align}

A particularly interesting case is when the set $\mathcal{N}$ corresponds to the set of all valid assemblages, \ie $\mathcal{N}=\{\pi_{a|x}|\sum_a \pi_{a|x}=\sum_a \pi_{a|x'}~\forall x,x',\tr\sum_a \pi_{a|0} = 1 \}$, in which case the quantifier was named simply the \emph{Steering Robustness $\SR(\cdot)$} \cite{PW15}. In this case $\tilde{\mathcal{N}}=\{\tilde{\pi}_{a|x}|\sum_a \tilde{\pi}_{a|x}=\sum_a \tilde{\pi}_{a|x'}~\forall x,x' \}$ and the first constraint in \eqref{eq: robustness final} is equal to $\sum_\lambda D(a|x,\lambda)\tilde{\sigma}_\lambda - \sigma_{a|x} \geq 0$. The dual formulation of \eqref{eq: robustness final} in this case is then given by
\begin{align}\label{eq: steering robustness dual}
\SR(\sigma_{a|x}) = \max_{\{F_{a|x}\}}& \quad \Tr \sum_{ax} F_{a|x}\sigma_{a|x}-1 \\
\text{s.t.}& \quad \openone - \sum_{ax} D_\lambda(a|x)F_{a|x} \geq 0 \quad\forall \lambda, \nonumber \\
	& \quad F_{a|x} \geq 0 \quad\forall a,x. \nonumber 
\end{align}
Similarly to the Steering Weight, the dual is seen to be strictly feasible, hence strong duality holds. We also see that the dual of the Steering Robustness has an interpretation in terms of the violation of steering inequalities such that $1\geq \tr\sum_{ax} F_{a|x}\sigma_{a|x}^\LHS$. In particular, it is equal to the maximisation over all positive linear steering functionals $\Tr \sum_{ax} F_{a|x}-1$ such that $1\geq \tr\sum_{ax} F_{a|x}\sigma_{a|x}^\LHS$. 

Other natural choices for the noise $\mathcal{N}$ are to consider (i) the set of LHS assemblages $\mathcal{N} = \{\pi_{a|x} | \pi_{a|x} = \sum_\lambda D(a|x,\lambda)\sigma_\lambda \, \forall a,x, \, \sigma_\lambda \geq 0 \, \forall \lambda, \tr\sum_\lambda \sigma_\lambda = 1\}$, corresponding to the \emph{LHS-Robustness} \cite{SAB+16}; (ii) composed by a single assemblage,  for example the maximally mixed assemblage, \ie $\mathcal{N} = \{\openone/(d_\rB o_\rA)~\forall a, x\}$\footnote{Remember that $o_\rA$ is the number of outputs of Alice's measurements, and $d_\rB$ is the dimension of Bob's Hilbert space}, corresponding to the \emph{Random Steering Robustness}. 

\subsection{Quantifying the steering of quantum states}
In the above we reviewed the question of how to quantify the steering of a given assemblage, \ie of the data observed in a given steering test. A related question is to quantify the steering of a quantum state. More precisely, one may want to optimise a given quantifier of steering over all possible steering tests performed on a given state, that is over any possible measurement strategy of Alice. More formally, for any given steering quantifier $\SQ(\cdot)$ of an assemblage $\sigma_{a|x}$, one can define
\begin{align}\label{e:SQ state}
\SQ(\rho^{\rA\rB}) = \max_{\{M_{a|x}\}} &\quad \SQ(\sigma_{a|x}) \\
\text{s.t.}&\quad \sigma_{a|x} = \tr_\rA[(M_{a|x}\otimes \openone)\rho^{\rA\rB}]\quad \forall~a,x,\nonumber 
\end{align}
where the first maximisation is understood to be over all possible sets of measurements, which in particular could include infinitely many measurements, with arbitrary numbers of outcomes\footnote{{In fact, for $d$-dimensional systems, considering POVMs with at most $d^2$ outcomes is sufficient. This follows since one can restrict to extremal POVMs due to the convexity of the underlying quantifier, and extremal POVMs have at most $d^2$ outcomes \cite{DPP05}.}}. If one nevertheless fixes the number of measurements $m_\rA$ and outcomes $o_\rA$, then this problem has the structure of the problem outlined in Appendix \ref{sec: see-saw}, and hence a see-saw algorithm can be employed to heuristically find measurement strategies which bound the given steering quantifier for a given state. In particular, given a steering inequality $\{F_{a|x}\}_{ax}$, and a fixed state $\rho^{\rA\rB}$, to find the optimal measurements is the following SDP
\begin{align}
\max_{\{M_{a|x}\}}&\quad \tr \sum_{ax} F_{a|x} \tr_\rA[(M_{a|x}\otimes \openone)\rho^{\rA\rB}] \\
\text{s.t.}&\quad \sum_a M_{a|x} = \openone \quad\forall x, \nonumber \\
&\quad M_{a|x} \geq 0 \quad\forall a,x. \nonumber 
\end{align}
This program, in conjunction with either \eqref{e:weight SDP dual final} or \eqref{eq: steering robustness dual} (or any other dual formulation of a steering quantifier) have the structure of \eqref{e:see-saw1} and \eqref{e:see-saw2}, and can thus be used in a see-saw algorithm. 

\subsection{Further results on steering quantification}

A few other steering quantifiers have been proposed so far besides the previously mentioned approach to quantify steering. In Ref.~\cite{GA15}, inspired by entanglement theory, the authors considered a more axiomatic approach to steering quantification. They first characterised the set of operations that cannot increase the amount of steering demonstrated by a given assemblage. Then, they defined steering monotones as those quantifiers that do not increase under these operations. Within this characterisation they proved that the steering weight and the steering robustness are proper monotones of steering and proposed the relative entropy of steering as a new monotone.

A different approach was followed in Refs.~\cite{KLRA15,KA15} to quantify the amount of steering in the continuous-variable scenario. There the authors employ the violation of a steering test based on the variances of certain observables as a building block. Moreover they show how to obtain experimentally-friendly lower bounds on this quantifier.

The quantification of steering in terms of the classical communication cost needed to simulate an assemblage was recently investigated in \cite{NV16,SAB+16}. In \cite{NV16} the set of assemblages which have an LHS model when $c$ bits of communication are sent from Alice to Bob was shown to have an SDP formulation. Using this they showed that the two-qubit Werner state has a model with one bit of communication for $w \leq 1/\sqrt{2}$. They also proved that infinite communication is necessary to simulate the maximally entangled state. In \cite{SAB+16} on the other hand, the LHS-robustness was shown to provide an upper bound on the amount of communication. Finally, they further showed that all pure entangled states, and some non-full rank states have infinite communication cost.

Finally, Refs.~\cite{CS16,CBLC16} showed how to bound robustness-based steering quantifiers in a fully device-independent way, \ie in a scenario that both Alice and Bob apply uncharacterised measurements.

\section{Quantum steering in multipartite scenarios}\label{Sec: multipartite}
In this section we consider the extension of quantum steering beyond the traditional bipartite setting, to the multipartite setting, consisting of several parties separated in space. There is some freedom in how exactly one generalises the bipartite steering scenario into a multipartite setting (see \cite{HR13,CSA+15}). In what follows we will focus on the particular approach taken in Ref.~\cite{CSA+15}, which is centred around the viewpoint of bipartite steering as an entanglement test where one of the parties performs uncharacterised measurements. At the end of this section we discuss other approaches one might take to generalise steering to the multipartite setting.

\begin{figure*}
\centering
\includegraphics[width=1.8\columnwidth]{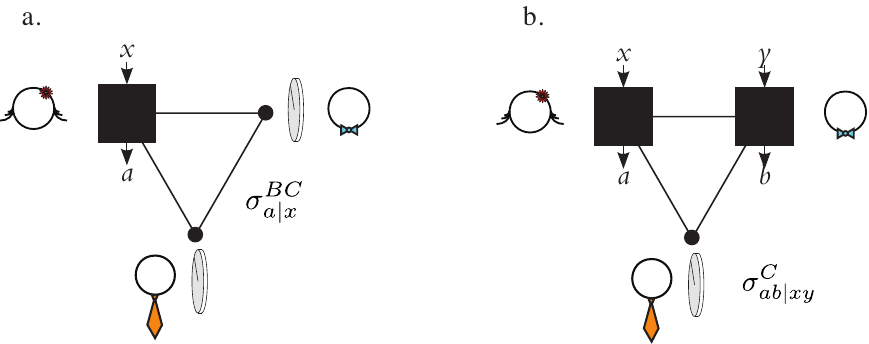}
\caption{Quantum steering in the tripartite case. a. One untrusted party: Alice measures her system with an uncharacterised measurement, leaving Bob and Charlie with the assemblage $\assem{\sigma}{a|x}{\rB\rC}$ \eqref{untrusted A}. b. Two untrusted parties: Alice and Bob apply uncharacterised measurements, leaving Charlie with the assemblage $\assem{\sigma}{ab|xy}{\rC}$ \eqref{untrusted AB}.}\label{f:scenariosTrip}
\end{figure*}

Consider as an example the tripartite scenario. There are two potential possibilities which one might consider as generalisations of the steering scenario (see Fig.~\ref{f:scenariosTrip}). In the first, one of the parties, say Alice, measures her system (which is assumed otherwise uncharacterised), and the bipartite states that are jointly prepared for Bob and Charlie can be analysed. That is, one can consider the assemblages
\be\label{untrusted A}
\assem{\sigma}{a|x}{\rB\rC}=\Tr_\rA[(M_{a|x}\otimes \openone^{\rB}\otimes\openone^{\rC})\rho^{\rA\rB\rC}].
\ee
Such a situation would then be the \emph{tripartite one-sided device-independent} scenario. In the second, both Alice and Bob measure their systems, and the states they jointly prepare for Charlie are analysed. That is, one considers the assemblages
\be\label{untrusted AB}
\assem{\sigma}{ab|xy}{\rC}=\Tr_{\rA\rB}[(M_{a|x}\otimes M_{b|y}\otimes \openone^\rC)\rho^{\rA\rB\rC}].
\ee
Accordingly, such a situation is the \emph{tripartite two-sided device-independent} scenario.

More generally, from this perspective, \emph{multipartite steering} scenarios consist of all the \emph{asymmetric network} scenarios, intermediate between the entanglement and Bell-nonlocality scenarios, where some subset of the parties have characterised devices, and the remaining subset uncharacterised devices. In each case there is a corresponding multipartite assemblage, similar to \eqref{untrusted A} and \eqref{untrusted AB}, which captures the observable data. As will become apparent below, taking this asymmetric network viewpoint as the definition of multipartite steering, effects more traditionally associated with entanglement and nonlocality will also come into play in a fundamental way.

In what follows we first review more formally the definition of multipartite steering as the detection of multipartite entanglement in an asymmetric quantum network and discuss how to treat this situation with SDP techniques. We will next show that the structure of tripartite steering is fundamentally richer than the bipartite case as the notion of post-quantum steering arises in this situation \cite{SBC+15}.

\subsection{Multipartite steering as the detection of multipartite entanglement in asymmetric networks}
In the remainder of this section we will discuss only the tripartite scenario in detail, since it captures all the ingredients present in the general multipartite scenario. We will discuss briefly in Sec.~\ref{s:arb multi} the additional technicalities that arise when considering more parties.

Recall from Sec.~\ref{Subsec: SDI ent} that bipartite steering can be seen as the certification of entanglement when Alice's measuring devices are uncharacterised. That is, separable states can only produce assemblages which possess the same structure as LHS assemblages, and therefore the observation of steering certifies that entanglement must have been shared between Alice and Bob.

Analogously, multipartite steering can be defined as the certification of entanglement when some subset of the parties have uncharacterised devices. The additional subtlety that arises is that in a multipartite setting there are multiple notions of separability, and accordingly of entanglement. In particular, the two extreme cases are (i) fully-separable states $\rho^\fs = \int d\lambda \mu(\lambda) \rho_\lambda^\rA\otimes \rho_\lambda^\rB \otimes \rho_\lambda^\rC$ and (ii) bi-separable states
\begin{align}\label{bisep rho}
\rho^\bs&=\int d\nu \mu(\nu)\rho^\rA_\nu\otimes\rho^{\rB\rC}_\nu+\int d\lambda \mu(\lambda)\rho^\rB_\lambda\otimes\rho^{\rA\rC}_\lambda\nonumber\\
&\quad+\int d\omega \mu(\omega)\rho^{\rA\rB}_\omega\otimes\rho^{\rC}_\omega,
\end{align}
\ie convex combination of states which are separable across at least one bipartition. States which are not fully separable are multipartite entangled and states which are not bi-separable are genuinely multipartite entangled (GME)\footnote{One can also consider asymmetric situations, for example states separable across the biparition A:BC, \ie those with the form of the first term on the right-hand-side of \eqref{bisep rho}. For the purpose of this review, we will not explicitly consider these situations.}. Having chosen a notion of separability/entanglement, and the number of uncharacterised parties, assemblages which arise from separable states will have additional structure that assemblages arising from entangled states do not necessarily have (just as in the bipartite case LHS assemblages \eqref{LHS} have additional structure that arbitrary assemblages \eqref{Born rule} do not necessarily have). Only when an assemblages does not have the additional structure is it said to demonstrate multipartite steering.

Let us start with the case of fully-separable states with one party using uncharacterised devices. This leads to assemblages of the form
\begin{align}\label{FS 1UN}
\assem{\sigma}{a|x}{\rB\rC}&=\Tr_\rA[(M_{a|x}\otimes \openone^{\rB}\otimes\openone^{\rC})\rho^\fs], \nonumber \\
&= \int d\lambda \mu(\lambda)p(a|x,\lambda) \rho_\lambda^\rB \otimes \rho_\lambda^\rC,
\end{align}
where $p(a|x,\lambda) = \tr[M_{a|x}\rho_\lambda^\rA]$. Thus, there is no steering between Alice and Bob-Charlie, and each member of the assemblage prepared is a separable state between Bob and Charlie. {Consider as a simple example the state $\ket{\phi^+}\bra{\phi^+}^{\rA\rB}\otimes\rho^\rC$. Since Alice and Bob share a steerable (maximally entangled) state, Alice can prepare an assemblage for Bob and Charlie that does not have the decomposition above, so Bob and Charlie can certify that they do not share a fully separable state. }

Alternatively, the above assemblage also arises from a multipartite LHS model, where with probability $\mu(\lambda)$ Alice receives the hidden variable $\lambda$ and outputs $a$ with probability $p(a|x,\lambda)$, Bob receives $\rho_\lambda^\rB$, and Charlie receives $\rho_\lambda^\rC$. 

Similarly, the case of fully-separable states when two parties use uncharacterised devices leads to assemblages of the form
\begin{align}\label{FS 2UN}
\assem{\sigma}{ab|xy}{\rC}&=\Tr_{\rA\rB}[(M_{a|x}\otimes M_{b|y}\otimes\openone^{\rC})\rho^\fs], \nonumber \\
&= \int d\lambda \mu(\lambda)p(a|x,\lambda)p(b|y,\lambda) \rho_\lambda^\rC,
\end{align}
where $p(b|y,\lambda) = \tr[M_{b|y}\rho_\lambda^\rB]$. Now, Alice and Bob share local correlations \citep{BCP+14}, and Alice-Bob are jointly unable to steer Charlie. Again,  {as a simple example, a state of the form $\rho^\rA\otimes\ket{\phi^+}\bra{\phi^+}^{\rB\rC}$ can be used to generate an assemblage that does not have this decomposition, simply because Bob and Charlie share a steerable state.}

Once more, there is a simple multipartite LHS model, identical to the above except that Bob now receives $\lambda$ instead of $\rho_\lambda^\rB$, and outputs $b$ with probability $p(b|y,\lambda)$.

The above suffices to characterise tripartite steering, and shares the closest analogue to the bipartite scenario, in the sense that LHS models which reproduce the assemblages that do not demonstrate multipartite steering are direct analogues of bipartite LHS models. Note also that when one party uses uncharacterised devices, it is crucial that the states prepared for Bob and Charlie are separable -- \ie elements of entanglement theory come in to the definition. Similarly, with two parties using uncharacterised devices it is crucial that the observed statistics between Alice and Bob are local -- \ie elements of nonlocality theory come into the definition.

Moving on to \emph{genuine tripartite steering}, the relevant notion of separability is now bi-separability \eqref{bisep rho}. The interest is now in the structure of assemblages which arise from measurements on bi-separable states (\ie those that do not demonstrate genuine multipartite steering). Starting with one uncharacterised device, the assemblages produced have the form
\begin{align}\label{GMS 1UN}
\assem{\sigma}{a|x}{\rB\rC}&=\Tr_\rA[(M_{a|x}\otimes \openone^{\rB}\otimes\openone^{\rC})\rho^\bs], \nonumber \\
&=\int d\nu \mu(\nu) p(a|x,\nu)\otimes\rho^{\rB\rC}_\nu+\int d\lambda \mu(\lambda)\rho^\rB_\lambda\otimes\assem{\sigma}{a|x,\lambda}{\rC}\nonumber\\
&\quad+\int d\omega \mu(\omega)\assem{\sigma}{a|x,\omega}{\rB}\otimes\rho^{\rC}_\omega,
\end{align}
where $\assem{\sigma}{a|x,\lambda}{\rC} := \tr_\rA[(M_{a|x}\otimes \openone^\rC)\rho_\lambda^{\rA\rC}]$ (and similarly for $\assem{\sigma}{a|x,\omega}{\rB}$). Such assemblages are seen to be a mixture of three terms, the salient features of which are (i) The first term does not demonstrate steering between Alice and Bob-Charlie. (ii) The second term consists only of separable states, and the marginal assemblage between Alice and Bob does not demonstrate steering\footnote{By marginal assemblage, we mean the assemblage that arises after tracing out the quantum states of some of the parties. That is, the marginal assemblage of Alice and Bob is $\assem{\sigma}{a|x}{\rB} = \tr_\rC[\assem{\sigma}{a|x}{\rB\rC}]$.}. (iii) The third term consists only of separable states and the marginal assemblage between Alice and Charlie does not demonstrate steering. 

When there are two parties using uncharacterised devices, the assemblages produced have the form
\begin{align}\label{GMS 2UN}
\assem{\sigma}{ab|xy}{\rC}&=\Tr_{\rA\rB}[(M_{a|x}\otimes M_{b|y}\otimes\openone^{\rC})\rho^\bs], \nonumber \\
&=\int d\nu \mu(\nu) p(a|x,\nu)\assem{\sigma}{b|y,\nu}{\rC}+\int d\lambda \mu(\lambda) p(b|y,\lambda)\assem{\sigma}{a|x,\lambda}{\rC}\nonumber\\
&\quad+\int d\omega \mu(\omega) p(ab|xy,\omega)\rho^{\rC}_\omega,
\end{align}
Again, such assemblages are mixtures of three terms, with the salient features (i) In the first term only Bob can steer the states of Charlie. (ii) In the second term, it is only Alice who can steer the states of Charlie. (iii) In the last term, Alice and Bob cannot jointly steer Charlie. They can however share \textit{quantum} nonlocal correlations between themselves.

Note that {the exemplary state given before, $\rho^\rA\otimes\ket{\phi^+}\bra{\phi^+}^{\rB\rC}$ will always produce assemblages like \eqref{GMS 1UN} or \eqref{GMS 2UN}, since it is a bi-separable state. What is more interesting is that there exist states that are genuinely multipartite entangled but that will always produce assemblages like these \cite{AAD+15}, or even like \eqref{FS 1UN} or \eqref{FS 2UN} \cite{TA06,BFF+16}.} Note that assemblages that do not demonstrate genuine multipartite steering can still demonstrate multipartite steering, which is a weaker notion.

{In what follows we will describe methods to detect these kinds of assemblages with SDP. While both notions of multipartite steering reduce to membership problems, \ie whether they have the above defined structure or not, they cannot be readily implemented with SDP.} In particular, there are two conditions which cannot be tested for in a necessary and sufficient way via semidefinite programming: (i) separability of members of the assemblage (ii) quantumness of the nonlocal behaviour.  However, in Appendix~\ref{Separability testing} and Sec.~\ref{quantumness testing} respectively, we address how each of these can be relaxed and dealt with through semidefinite programming. The former is based upon the \emph{$k$-extendibility hierarchy} of Doherty, Parillo and Spedalieri \cite{DPS02}, while the latter is based upon the Navascu\'es, Pironio and Ac\'in hierarchy \cite{NPA08}. Using these techniques, SDPs can be utilised to test for a necessary set of conditions -- that is, it can be used to define an \emph{outer approximation} to the set of assemblages which do not demonstrate steering. If an assemblage is found not to satisfy the necessary conditions, in other words lie outside the outer approximation), then one certifies that steering has been demonstrated.

\subsection{Quantumness testing}\label{quantumness testing}

In \eqref{GMS 2UN}, in the last term on the right-hand-side, the assemblage has the property that Alice and Bob can observe nonlocal correlations between the outcomes of their uncharacterised devices. These correlations should nevertheless be quantum, since the underlying assumption was that they arose from measurements on a quantum state. Deciding if a nonlocal behaviour has a quantum description or not turns out to be a hard problem. Navascu\'es, Pironio and Ac\'in (NPA) introduced a hierarchy of SDPs, based upon moment matrices, which tests for membership in the set of quantum nonlocal behaviours. This hierarchy converges to the quantum set in the limit \cite{NPA08}.

The NPA hierarchy of moment matrices can be generalised to the case of steering, such that it provides a hierarchy of tests which an assemblage must pass if it is to have a quantum realisation. This idea, based upon a bipartite-modification of the NPA hierarchy \cite{MBL+13} was introduced in \cite{NTV14,P13} (see also \cite{CSA+15,SBC+15,JMRW15}).

Consider the set $\mathcal{M} = \{M_{a|x} \}_{a,x} \cup \{M_{b|y}\}_{b,y}$ of all measurement operators, which can without loss of generality be taken to be projective, $M_{a|x} M_{a'|x} = \delta_{a,a'}M_{a|x}$ (and similarly for Bob)\footnote{If the operators are POVM elements and not projectors, we extend to a larger Hilbert space, such that they are indeed projectors, and consider this extension instead.}. Starting from this set, the set of all strings of operators of length $k$ or less can be formed,
\begin{align}
\mathcal{S}_0 &= \{\openone\otimes \openone\}, \nonumber \\
\mathcal{S}_k &= \mathcal{S}_{k-1} \cup \{M_i S_j | S_j \in \mathcal{S}_{k-1}, M_i \in \mathcal{M}\}.
\end{align}
Note that, due to the orthogonality relation between the elements of projective measurements, some elements of $\mathcal{S}_k$ vanish, and these are assumed to be implicitly omitted in the above definition. Now, to each set $\mathcal{S}_k$ we associate a completely positive (CP) map $\Lambda^{(k)}$, taking operators acting on $\mathcal{H}_\rA \otimes \mathcal{H}_\rB$ to operators acting on $\mathcal{H}_\mathrm{O} = \mathbb{C}^{d_k}$, where $d_k = |\mathcal{S}_k|$, with Kraus operators $A_n = \sum_i \ket{i}\bra{n}S_i$, where $\{\ket{i}\}$ forms a basis for $\mathcal{H}_\mathrm{O}$, and $\{\ket{n}\}$ a basis for $\mathcal{H}_\rA \otimes \mathcal{H}_\rB$. The action of this CP map, applied to the tripartite state $\rho^{\rA\rB\rC}$ is
\begin{equation}\label{NPA channel}
\Lambda^{(k)}\otimes \mathrm{id} (\rho^{\rA\rB\rC}) = \sum_{i,j}\ket{i}\bra{j} \otimes \tr_{\rA\rB}[(S_j^\dagger S_i \otimes \openone^\rC) \rho^{\rA\rB\rC}].
\end{equation}
As a CP map, the output must be a valid quantum state, and in particular positive-semidefinite. Now, certain parts of the output contain the multipartite assemblage $\assem{\sigma}{ab|xy}{\rC}$. That is, whenever $S_j^\dagger S_i$ (or a linear combination thereof) can be expressed as a linear combination of at most one measurement operator for each Alice and Bob, namely if
\begin{multline}\label{e:NPA consist}
\sum_{ij}\xi_{ij} S_j^\dagger S_i = \alpha (\openone\otimes\openone) + \sum_{ax}\beta_{ax} (M_{a|x} \otimes \openone) \\+ \sum_{by}\gamma_{by} (\openone \otimes M_{b|y}) + \sum_{abxy} \delta_{abxy}(M_{a|x}\otimes M_{b|y}),
\end{multline}
for some coefficients $\{\xi_{ij}, \alpha, \beta_{ax}, \gamma_{by}, \delta_{abxy}\}$, then, by defining $\Xi := \sum_{ij}\xi_{ij}\ket{j}\bra{i}$, 
\begin{equation}
\tr_\mathrm{O}\Big[(\Xi \otimes \openone) \Lambda^{(k)}\otimes \mathrm{id} (\rho^{\rA\rB\rC})\Big] = \Theta(\assem{\sigma}{ab|xy}{\rC}),
\end{equation}
where
\begin{multline}
\Theta(\assem{\sigma}{ab|xy}{\rC}) := \alpha\rho^\rC
+ \sum_{ax}\beta_{ax}\assem{\sigma}{a|x}{\rC}\\
+ \sum_{by}\gamma_{by}\assem{\sigma}{b|y}{\rC} + \sum_{abxy}\delta_{abxy}\assem{\sigma}{ab|xy}{\rC},
\end{multline}
and where $\assem{\sigma}{a|x}{\rC} := \sum_b \assem{\sigma}{ab|xy}{\rC}$ (which is independent of $y$ due to no-signalling), $\assem{\sigma}{b|y}{\rC} := \sum_a \assem{\sigma}{ab|xy}{\rC}$ and $\rho^\rC := \sum_{ab} \assem{\sigma}{ab|xy}{\rC}$.

Thus, for a given assemblage, a necessary condition for it to have a quantum realisation is that it is consistent with being the partial output of a CP map of the form \eqref{NPA channel}. Since the consistency requirements are linear in the elements of the assemblage, this thus becomes a feasibility SDP:
\begin{align}\label{e:NPA feas}
\text{given}& \quad\{\assem{\sigma}{ab|xy}{\rC}\}_{abxy}, \,k \nonumber \\
\text{find}& \quad \Gamma^{(k)} \\
\text{s.t.}& \quad \tr_\mathrm{O}\Big[(\Xi_{i}\otimes \openone) \Gamma^{(k)}\Big] = \Theta_i(\assem{\sigma}{ab|xy}{\rC}), \nonumber \\
&\quad \Gamma^{(k)} \geq 0, \nonumber
\end{align}
where $\Gamma^{(k)}$ is the $d_kd_\rC \times d_kd_\rC$ output of the CP map $\Lambda^{(k)}\otimes \mathrm{id}(\cdot)$, and the consistency requirements are understood to hold for all independent sets $\{\Xi_{i}, \Theta_i(\cdot)\}$ which arise from \eqref{e:NPA consist}. If no such $\Gamma^{(k)}$ can be found, this implies that the assemblage does not have a quantum realisation. We will use the notation $\{\assem{\sigma}{ab|xy}{\rC}\} \in \mathcal{Q}^\rC_k$ if an assemblage is  contained in level $k$ of the steering-NPA hierarchy, \ie if it satisfies the above feasibility SDP. In \cite{JMRW15} it was shown that as $k\to \infty$, like the original NPA hierarchy, the steering-NPA hierarchy converges.

\subsection{SDP tests for multipartite and genuine-multipartite steering}
Given the SDP approximations for separability, outlined in Appendix~\ref{Separability testing}, and for quantumness, outlined in Sec.~\ref{quantumness testing}, the final SDP tests for whether a multipartite assemblage demonstrates multipartite steering or genuine multipartite steering can be given, when either one or two parties have uncharacterised devices. Starting with the former case of multiparite steering, the feasibility SDP for one uncharacterised device is:
\begin{align}\label{e:mult steering 1UN}
\text{given}& \quad\{\assem{\sigma}{a|x}{\rB\rC}\}_{ax},\, k \nonumber \\
\text{find}& \quad \{\sigma_\lambda^{\rB\rC}\}_\lambda \\
\text{s.t.}& \quad \sum_\lambda D(a|x,\lambda)\sigma_\lambda^{\rB\rC} = \assem{\sigma}{a|x}{\rB\rC}\quad \forall a,x, \nonumber \\
&\quad \{\assem{\sigma}{\lambda}{\rB\rC}\} \in \assem{\Sigma}{k-\mathrm{sep}}{\rB\rC}. \nonumber
\end{align}
where $\{\assem{\sigma}{\lambda}{\rB\rC}\} \in \assem{\Sigma}{k-\mathrm{sep}}{\rB\rC}$ denotes that each member of the LHS model should have a (unnormalised) $k$-symmetric extension, as outlined in Appendix~\ref{Separability testing}. Note that we have used the fact here that $\{\assem{\sigma}{\lambda}{\rB\rC}\}_\lambda$ can be understood as an assemblage, where Alice makes only a single measurement, with outcome $\lambda$. 

If an assemblage passes this test for a given $k$, then no conclusion can be drawn. If on the other hand an assemblage fails the test, then it certifies that it demonstrates multipartite steering, as it is incompatible with the structure that any assemblage arising from a fully-separable state necessarily has. For two uncharacterised devices, the test is
\begin{align}\label{e:mult steering 2UN}
\text{given}& \quad\{\assem{\sigma}{ab|xy}{\rC}\}_{abxy} \nonumber \\
\text{find}& \quad \{\sigma_{\mu\nu}^{\rC}\}_{\mu\nu} \\
\text{s.t.}& \quad \sum_{\mu\nu} D(a|x,\mu)D(b|y,\nu)\sigma_{\mu\nu}^{\rC} = \assem{\sigma}{ab|xy}{\rC} \quad \forall a,b,x,y, \nonumber \\
&\quad \assem{\sigma}{\mu\nu}{\rC} \geq 0 \quad \forall \mu, \nu. \nonumber
\end{align}

For the case of genuine multipartite steering, the feasibility SDP for one uncharacterised device takes the form
\begin{align}\label{e:gen mult steering 1UN}
\text{given}& \quad\{\assem{\sigma}{a|x}{\rB\rC}\}_{ax}, \,k \nonumber \\
\text{find}& \quad \{\assem{\sigma}{\mu}{\rB\rC}\}_\mu,\, \{\assem{\pi}{a|x}{\rB\rC}\}_{ax},\, \{\assem{\pi}{\nu}{\rC}\}_\nu,\, \{\assem{\gamma}{a|x}{\rB\rC}\}_{ax},\, \{\assem{\gamma}{\lambda}{\rB}\}_\lambda \\
\text{s.t.}& \quad \sum_\mu D(a|x,\mu)\assem{\sigma}{\mu}{\rB\rC} + \assem{\pi}{a|x}{\rB\rC} + \assem{\gamma}{a|x}{\rB\rC}  = \assem{\sigma}{a|x}{\rB\rC}\quad \forall a,x, \nonumber \\
&\quad\tr_\rB[\assem{\pi}{a|x}{\rB\rC}] =\sum_\nu D(a|x,\nu)\assem{\pi}{\nu}{\rC}\quad \forall a,x, \nonumber \\
&\quad\tr_\rC[\assem{\gamma}{a|x}{\rB\rC}] =\sum_\lambda D(a|x,\lambda)\assem{\gamma}{\lambda}{\rB}\quad \forall a,x, \nonumber \\
&\quad \{\assem{\pi}{a|x}{\rB\rC}\} \in \assem{\Sigma}{k-\mathrm{sep}}{\rB\rC},\quad \{\assem{\gamma}{a|x}{\rB\rC}\} \in \assem{\Sigma}{k-\mathrm{sep}}{\rB\rC},\nonumber \\
&\quad \assem{\sigma}{\mu}{\rB\rC}\geq 0\quad \forall \mu, \quad \assem{\pi}{\nu}{\rC} \geq 0\quad \forall \nu, \quad \assem{\gamma}{\lambda}{\rB} \geq 0\quad \forall \lambda. \nonumber
\end{align}
If the above problem is infeasible for a given $k$, it certifies that the assemblage $\assem{\sigma}{a|x}{\rB\rC}$ is inconsistent within having come from a biseparable state, and thus demonstrates genuine multipartite steering. Finally, with two uncharacterised devices, the feasibility SDP is
\begin{align}\label{e:gen mult steering 2UN}
\text{given}& \quad\{\assem{\sigma}{ab|xy}{\rC}\}_{abxy}, \,k \nonumber \\
\text{find}& \quad \{\assem{\pi}{b|y,\mu}{\rC}\}_{by\mu},\, \{\assem{\pi}{a|x,\nu}{\rC}\}_{ax\nu},\, \{\assem{\pi}{\lambda}{\rC}\}_\lambda \\
\text{s.t.}& \quad \sum_\mu D(a|x,\mu)\assem{\pi}{b|y,\mu}{\rC} + \sum_\nu D(b|y,\nu)\assem{\pi}{a|x,\nu}{\rC} \nonumber \\
&\quad\quad + \sum_\lambda D_\mathrm{NS}(ab|xy,\lambda)\assem{\pi}{\lambda}{\rC} = \assem{\sigma}{ab|xy}{\rC}\quad \forall a,b,x,y,\nonumber \\
&\quad\Big\{\sum_\lambda D_\mathrm{NS}(ab|xy,\lambda)\assem{\pi}{\lambda}{\rC}\Big\} \in \mathcal{Q}^\rC_k. \nonumber
\end{align}
where $D_\mathrm{NS}(ab|xy,\lambda)$ denotes the vertices of the non-signalling polytope \cite{BCP+14}, and where for each $\mu$, $\{\assem{\pi}{b|y,\mu}{\rC}\}_{b,y}$ is understood to be a valid sub-normalised assemblage (\ie such that $\assem{\pi}{b|y,\mu}{\rC} \geq 0$ $\forall b,y,\mu$, $\sum_b \assem{\pi}{b|y,\mu}{\rC} = \sum_b \assem{\pi}{b|y',\mu}{\rC}$ $\forall y,y'\mu$ but $\tr[\sum_b \assem{\pi}{b|y,\mu}{\rC}] \leq 1$), and similarly for $\{\assem{\pi}{a|x\,\nu}{\rC}\}_{a,x}$. Again, if for a given $k$ the above SDP is infeasible, it certifies that the assemblage $\assem{\sigma}{ab|xy}{\rC}$ is incompatible with having come from a bi-separable state, and thus demonstrates genuine multipartite steering.

\subsubsection{Examples}
Two exemplary genuine-multipartite-entangled states are the W and GHZ states, $\ket{W} = (\ket{001} + \ket{010} + \ket{100})/\sqrt{3}$ and $\ket{GHZ} = (\ket{000} + \ket{111})/\sqrt{2}$. In \cite{CSA+15} the critical robustness to white noise for the demonstration of multipartite and genuine multipartite steering of these states were studied. {The authors found upper bounds }on the maximum $w$ such that
\be\label{eq: noisy tripartite}
w \ket{\psi}\bra{\psi} + (1-w)\openone/8
\ee
 does not demonstrate multipartite or genuine multipartite steering, where $\ket{\psi} = \ket{W}$ or $\ket{\mathrm{GHZ}}$. The results can be found in Table~\ref{examples GHZ W}.
 
\begin{table}[h]
\centering
\begin{tabular}{c|cc|c|c}
 & \multicolumn{2}{c|}{GHZ} & \multicolumn{2}{c}{W} \\ \cline{2-5} 
 & \multicolumn{1}{c|}{MS} & \multicolumn{1}{c|}{GMS} & \multicolumn{1}{c|}{MS} & \multicolumn{1}{c}{GMS} \\ \hline
1 uncharacterised device & \multicolumn{1}{c|}{\qu 0.2500\qu } & \qu 0.5420\qu  & \qu 0.2698\qu  & \qu 0.5684\qu  \\ \hline
2 uncharacterised devices & \multicolumn{1}{c|}{\qu 0.4286\qu } & \qu 0.6322\qu  & \qu 0.4434\qu  & \qu 0.6757\qu 
\end{tabular}%
\caption{{Upper} bound on the critical robustness to white noise for the tripartite GHZ and W states to demonstrate multipartite steering (MS) or genuine multipartite steering (GMS), when three measurements are performed by the (one or two) parties using uncharacterised devices. }
\label{examples GHZ W}
\end{table}
 
\subsection{Multipartite steering inequalities}
In the previous section we reviewed SDP tests which certify whether a given multipartite assemblage demonstrates some form of multipartite steering or not. Just as in the bipartite case, through SDP duality, it is possible to write down a dual formulation of each of the above programs which has an interpretation in terms of \emph{multipartite steering inequalities}. In particular, for the case of one uncharacterised device, the dual has the generic form
\begin{align}
\text{given} &\quad \{\assem{\sigma}{a|x}{\rB\rC}\}_{ax} \nonumber \\
\min_{\{F_{a|x}\}} &\quad \tr\sum F_{a|x}\assem{\sigma}{a|x}{\rB\rC} \\
\text{s.t.}&\quad \{F_{a|x}\} \in \mathcal{F} \nonumber 
\end{align}
where $\{F_{a|x}\} \in \mathcal{F}$ signifies that the inequality operators $F_{a|x}$ must satisfy a number of PSD constraints and/or LMIs, depending on the particular type of entangled tested for. These constraints will ensure that $\sum_{ax} F_{a|x}\assem{\pi}{a|x}{\rB\rC} \geq 0$ for all multipartite assemblages $\assem{\pi}{a|x}{\rB\rC}$ which do not demonstrate the particular type of steering. Thus a value $\beta =  \tr\sum F_{a|x}\assem{\sigma}{a|x}{\rB\rC} < 0$ certifies the demonstration of multipartite steering. 

For the case of two uncharacterised devices, the generic form of the dual is similarly
\begin{align}
\text{given} &\quad \{\assem{\sigma}{ab|xy}{\rC}\}_{abxy} \nonumber \\
\min_{\{F_{ab|xy}\}} &\quad \tr\sum F_{ab|xy}\assem{\sigma}{ab|xy}{\rC} \\
\text{s.t.}&\quad \{F_{ab|xy}\} \in \mathcal{F} \nonumber 
\end{align}
where again $\mathcal{F}$ depends on the type of entanglement tested for, and ensures that $\tr\sum F_{ab|xy}\assem{\pi}{ab|xy}{\rC} \geq 0$ holds for all assemblages $\assem{\pi}{ab|xy}{\rC}$ that to not demonstrate the particular type of multipartite steering tested for. 
\subsection{Generalisations to more parties}\label{s:arb multi}
We have reviewed so far only the case of tripartite steering. Notice however that the above can readily be extended to derive SDPs to test the presence of different kinds of entanglement for general $N$-partite systems.

First of all one specifies the scenarios by fixing (i) a particular type of entanglement and (ii) the pattern of which parties use characterised or uncharacterised devices. The entanglement can be chosen arbitrarily, for example one may ask that the state is not fully separable, be separable across a given number of fixed bipartitions, or be a convex combination of states separable over a given number of partitions (but not necessarily fixed). The pattern may also be chosen arbitrarily, ranging from all but one party using characterised devices, to all but one using uncharacterised devices.

Given the specification, one then enumerates the list of properties which the corresponding multipartite assemblages have. These properties will fall into two classes: those which impose constraints which are directly applicable, \ie are in the form of PSD constraints and LMIs, and those which are not. As in the tripartite case, the approach is then to relax the non-directly applicable constraints to find an approximate SDP test.

The main drawback in this approach is that as the number of parties increases, the difficulty of the problem will grow to the point where numerical techniques are unable to solve efficiently the tests. For example, one class of constraints that will arise is that multipartite assemblages will need to have quantum realisations. In principle such a constraint can still be imposed by extending the steering NPA hierarchy to deal with arbitrary numbers of uncharacterised devices. However, in the multipartite setting this soon becomes intractable. Alternatively, one may have constraints that a multipartite quantum assemblage is formed only of separable states. One can again relax this using the multipartite generalisation \cite{DPS05} of the $k$-extendibility techniques  presented in Appendix~\ref{Separability testing}.

In summary, the approach presented above is most suitable to scenarios involving relatively small numbers of parties, where it provides powerful tests for multipartite entanglement (and explicitly provides witnesses in each case). This is similar to the situation for multipartite entanglement (where the increase of the Hilbert space dimension limits numerical techniques), and in the fully device-independent approach (where the growth of the dimension of the space of correlations, and the number of nonlocal vertices, limits techniques).
\subsection{Post-quantum steering}
In the previous section on the detection of tripartite entanglement with two uncharacterised devices, one of the subtleties that arose was that it is not automatic that an assemblage $\assem{\sigma}{ab|xy}{\rC}$ has a quantum realisation, \ie that it can be generated by some state $\rho^{\rA\rB\rC}$ and some measurements for Alice and Bob, $M_{a|x}$ and $M_{b|y}$, through Eq.~\eqref{untrusted AB}. In particular, the minimal possible physical requirements that any assemblage must satisfy is \emph{no-signalling} -- that no party should be able to instantaneously communicate to any other group of parties. For multipartite assemblages, this is guaranteed if
\begin{align}\label{eq:non-signalling}
\sum_a \assem{\sigma}{ab|xy}{\rC} &= \sum_a \assem{\sigma}{ab|x'y}{\rC}\quad \forall x,x',b,y. \nonumber \\
\sum_b \assem{\sigma}{ab|xy}{\rC} &= \sum_b \assem{\sigma}{ab|xy'}{\rC}\quad \forall a, x,y,y'.
\end{align}
However, in the tripartite case, not every non-signalling assemblage necessarily arises from measurements on a quantum state. Indeed, consider an assemblage of the form $\assem{\sigma}{ab|xy}{\rC} = p_\mathrm{NS}(ab|xy)\rho^\rC$, where $p_\mathrm{NS}(ab|xy)$ is any non-signalling behaviour which cannot be realised in quantum theory (for example, the Popescu-Rohrlich box \cite{PR94}), and $\rho^\rC$ is an arbitrary quantum state. Clearly this multipartite assemblage satisfies the non-signalling conditions \eqref{eq:non-signalling}, yet it could never been produced in quantum theory, since if it could then it would provide a means to produce post-quantum behaviours within quantum theory, which is a contradiction. What this shows is that trivially the existence of post-quantum nonlocality implies the existence of post-quantum steering. This was why the set $\mathcal{Q}^\rC_k$ was introduced in Sec.~\ref{quantumness testing}, in order to approximate those assemblages that arise in quantum theory.

In \cite{SBC+15} the question was explored as to whether there is more interest to post-quantum steering than the above simple example. That is, the above relied only on the existence of post-quantum nonlocality, and thus does not imply that post-quantum steering is a genuinely different phenomenon from it. Given the amount of interest there has been in post-quantum nonlocality \cite{BCP+14}, and the successes this has lead to in understanding the structure of quantum nonlocality, it is relevant to know if steering allows one to go beyond quantum theory in ways inequivalent to how nonlocality does.

To show that post-quantum steering is a genuinely different phenomenon from post-quantum nonlocality, it is necessary to find a post-quantum assemblage that is only able to produce nonlocal behaviours which have a quantum realisation. That is, the pair of conditions:
\begin{subequations}
\begin{align}
\assem{\sigma}{ab|xy}{\rC} &\neq \Tr_{\rA\rB}[(M_{a|x}\otimes M_{b|y}\otimes \openone^\rC)\rho^{\rA\rB\rC}], \label{e:nonquantum}\\
\tr[M_{c|z}\assem{\sigma}{ab|xy}{\rC}] &= \Tr[(M'_{a|x}\otimes M'_{b|y}\otimes M'_{c|z})\rho'^{\rA\rB\rC}] \label{e:quantum all meas}.
\end{align}
\end{subequations}
where the first line says that it is impossible to find a triple $\{M_{a|x}, M_{b|y}, \rho^{\rA\rB\rC}\}$ that reproduce the assemblage, and the second says that for any set of measurements (in particular even for all POVMs), there should be a quantum description $\{M'_{a|x}, M'_{b|y}, M'_{c|z}, \rho'^{\rA\rB\rC}\}$ of the behaviour $p(abc|xyz) := \tr[M_{c|z}\assem{\sigma}{ab|xy}{\rC}]$.

These two conditions can be efficiently enforced using SDP techniques for the case where the Charlie holds qutrits, \ie $\mathcal{H}_\rC = \mathbb{C}^3$, as we review in the following.

 The first condition can be checked by using the analogue of Tsirelson bounds for multipartite steering inequalities. In particular, given a steering functional with operators $F_{ab|xy}$, one can upper bound the maximal value obtained by any quantum realisable assemblage by using the steering-NPA hierarchy introduced in Sec~\ref{quantumness testing}. In particular, one can evaluate the SDP
\begin{align}
\text{given}&\quad \{F_{ab|xy}\}_{abxy}\nonumber \\
\beta^\mathcal{Q}_k = \max_{\{\assem{\pi}{ab|xy}{\rC}\}}& \tr\sum_{abxy} F_{ab|xy} \assem{\pi}{ab|xy}{\rC} \\
\text{s.t.}&\quad \{\assem{\pi}{ab|xy}{\rC}\} \in \mathcal{Q}_k^\rC \nonumber
\end{align}
Thus, if the assemblage $\assem{\sigma}{ab|xy}{\rC}$ achieves the value $\beta = \tr \sum_{abxy}F_{ab|xy}\assem{\sigma}{ab|xy}{\rC} > \beta_k^\mathcal{Q}$, then this certifies that condition \eqref{e:nonquantum} holds, \ie that the assemblage is post-quantum.

The second condition can be enforced by first using a variant of the method outlined in Sec.~\ref{sec:infinite to finite} for passing from all projective measurements to a finite set of measurements for qubits, and then by passing from all projective measurements on qubits to all POVM measurements on qutrits. In particular, a stronger condition than \eqref{e:quantum all meas} is that $\tr[M_{c|z}\assem{\sigma}{ab|xy}{\rC}]$ is \emph{local} for the measurements $M_{c|z}$. Using the method of Sec.~\ref{sec:infinite to finite}, if $M_{c|z}$ corresponds to all qubit projective measurements, this is equivalent to $\tr[\Pi_{c|z}\assem{\gamma}{ab|xy}{\rC}]$ being local for a finite set of projective measurements $\{M_z\}_z$, with POVM elements $\Pi_{c|z}$ and largest inscribed ball of radius $r$, where $\assem{\sigma}{ab|xy}{\rC} = r \assem{\gamma}{ab|xy}{\rC} + (1-r)\tr[\assem{\gamma}{ab|xy}{\rC}]\openone/2$. Finally, from the results of \cite{HQB+13}, if the qubit assemblage $\assem{\sigma}{ab|xy}{\rC}$ produces local behaviours for all projective measurements on Charlie, then the qutrit assemblage $\assem{\tilde{\sigma}}{ab|xy}{\rC} := \tfrac{1}{3} \assem{\sigma}{ab|xy}{\rC} + \tfrac{2}{3}\tr[\assem{\sigma}{ab|xy}{\rC}]\proj{2}$ produces local behaviours for all POVM measurements, and therefore satisfies condition \eqref{e:quantum all meas}. Putting everything together, the following SDP problem can be used to test for post-quantum steering:
\begin{align}
\text{given} &\quad \{F_{ab|xy}\}_{abxy}, \{\Pi_{c|z}\}_{cz}, r \nonumber \\
\max_{\{\gamma_{ab|xy}\}, \{q_{\mu\nu\lambda}\}}\!\! &\quad \tr\sum_{abxy}F_{ab|xy}\assem{\tilde{\sigma}}{ab|xy}{\rC} \\
\text{s.t.} &\quad \assem{\tilde{\sigma}}{ab|xy}{\rC} = \tfrac{1}{3}[r \assem{\gamma}{ab|xy}{\rC} + \tr[\assem{\gamma}{ab|xy}{\rC}]((1-r)\tfrac{\openone}{2} + 2\proj{2})] \nonumber\\
&\quad \tr[\Pi_{c|z}\assem{\gamma}{ab|xy}{\rC}] = \sum_{\mu\nu\lambda}q(\mu\nu\lambda) D(abc|xyz,\mu\nu\lambda) \nonumber \\
&\quad \sum_a \assem{\gamma}{ab|xy}{\rC} = \sum_a \assem{\gamma}{ab|x'y}{\rC},\quad
\sum_b \assem{\gamma}{ab|xy}{\rC} = \sum_b \assem{\gamma}{ab|xy'}{\rC} \nonumber \\
&\quad\tr\sum_{ab} \assem{\gamma}{ab|xy}{\rC} = 1, \quad \assem{\tilde{\sigma}}{ab|xy}{\rC} \geq 0 \nonumber \\
&\quad\sum_{\mu\nu\lambda} q(\mu\nu\lambda) = 1, \quad q(\mu\nu\lambda) \geq 0 \nonumber
\end{align} 
where $D(abc|xyz,\mu\nu\lambda) \equiv D(a|x,\mu)D(b|y,\mu)D(c|z,\lambda)$ is a (product) local deterministic behaviour for Alice, Bob and Charlie. 

By appropriately searching over inequalities $F_{ab|xy}$, one can thus find examples of assemblages which display post-quantum steering, yet which produce only local behaviours, as was demonstrated in \cite{SBC+15}. Thus, there is a form of post-quantum steering which is inequivalent to post-quantum nonlocality. 

\subsection{Further results on multipartite steering}\label{sec: other multipartite}

A different approach to multipartite steering was proposed by He and Reid \cite{HR13}. Motivated by the idea of an LHS model in the bipartite case, they define the detection of genuine multipartite steering when a probability distribution does not satisfy the following model (restricting to the tripartite case):
\begin{align}
&\sum_{\lambda}p(\lambda)p_Q(ab|xy,\lambda)p(c|z,\lambda)\nonumber\\
&+\sum_{\nu}p(\nu)p_Q(ac|xz,\nu)p(b|y,\nu)\nonumber\\
&+\sum_{\mu}p(\mu)p(a|x,\mu)p_Q(bc|yz,\mu),
\end{align}
where the label $Q$ indicates that the corresponding probability distribution comes from specific (known) measurements applied by that party. Notice that a key difference with the previous approach is that now Alice, Bob and Charlie are all assumed to perform well characterised measurements at some point in the decomposition (\ie who is characterised and who is not is not fixed). Such a definition of genuine multipartite steering was recently demonstrated experimentally in Refs. \cite{AWT+15,LCC+15}.

Finally, the idea of the existence of post quantum steering in the tripartite section was, to the best of our knowledge,  first noticed in \cite{F12}. Steering in the context of generalised probabilistic theories, (which is a different notion of post quantum steering than that introduced here, since there no party holds a quantum state) has also been investigated in the bipartite case in Ref.~\cite{B15}.

\section{Continuous variable systems}
\label{sec: cv systems}
In everything covered up until this point, we have reviewed only the case of finite dimensional Hilbert spaces. In particular, it was assumed that the members of the assemblage $\sigma_{a|x}$ are finite dimensional -- \ie that Bob's Hilbert space is finite dimensional. Moreover, whenever we have reviewed questions about the steerability of specific quantum states $\rho^{\rA\rB}$, we have only considered the case that Alice's Hilbert space is also finite dimensional. 

One can however consider situations where the Hilbert spaces under consideration are infinite dimensional -- \ie continuous variable (CV) systems. For instance, one may be interested in situations where Alice's measurements have continuous outcomes, the members of the assemblage are infinite dimensional, or the question of which multipartite continuous variable systems are steerable.

Clearly, the SDP methods presented so far are not suitable for these cases. 
In what follows, we will review how the steering of CV systems can still be investigated with SDP, using the notion of a \emph{moment matrix} of expectation values  \cite{KSC+15}. 
This idea is closely related to the the NPA hierarchy of moment matrices used in nonlocality \cite{NPA08} (and the associated steering-NPA hierarchy discussed in Sec.~\ref{quantumness testing}) and the moment matrix method for entanglement detection of CV states \cite{SV05}. 

\subsection{Moment matrix approach to steering} \label{Sec: moment}

The starting point of the moment matrix is first to introduce the moments themselves. As previously, we assume that Alice and Bob share a state $\rho^{\rA\rB}$, and that Alice performs measurements $M_{a|x}$, where $a$ is now a continuous outcome. Note that these measurements can be taken to be ideal projective rank-1 measurements without loss of generality -- we can always extend the dimension of Alice's Hilbert space such that this is the case, due to Naimark's theorem. 

Formally, by performing such measurements, Alice still produces an assemblage of the form \eqref{Born rule} for Bob. However, the goal is to avoid using this directly due to the difficulties of it containing infinitely many members, and infinite dimensional states. Therefore it is replaced by two physically accessible pieces of information. First, Bob is also taken to perform ideal projective measurements $M_{b|y}$, for $y = 0,\ldots, m_\rB-1$, which have associated observables $B_y := \int \mathrm{d}b~ b M_{b|y}$, and look at the moments 
\begin{align}\label{e:moments}
\expect{A_x^m B_y^n} &:= \iint \mathrm{d}a ~\mathrm{d}b~ a^m b^n \tr[(M_{a|x}\otimes M_{b|y})\rho^{\rA\rB}], \nonumber \\
&= \int \mathrm{d}a~ a^m \tr[\sigma_{a|x} B_y^n], \nonumber \\
&= \iint \mathrm{d}a~ \mathrm{d}b~ a^m b^n P(ab|x,B_y),
\end{align} 
which can be estimated by Alice and Bob. 

Furthermore, Bob will be allowed to estimate any operator on his system, \ie for any operator $B$, it is assumed that Bob can estimate $\expect{B} := \tr[B\rho^\rB]$\footnote{This could be done, for example, by first estimating the Wigner function, and then calculating by hand the expectation of $B$ from the Wigner function}. Note that these assumptions reflect, and maintain, the asymmetry of the steering scenario -- Bob knows precisely the measurements he is performing, whilst no assumption is made about Alice, including the dimension of her Hilbert space. 

Now, much like in Sec.~\ref{quantumness testing}, one can consider the set of observables $\mathcal{O} = \{A_x\}_x\cup \{B_y\}_y$, where $A_x := \int \mathrm{d}a ~a M_{a|x}$, and the sets
\begin{align}
\mathcal{T}_0 &= \{\openone\otimes \openone\}, \nonumber \\
\mathcal{T}_k &= \mathcal{T}_{k-1} \cup \{O_i T_j | T_j \in \mathcal{T}_{k-1}, O_i \in \mathcal{O}\}.
\end{align}

To each set $\mathcal{T}_k$ is associated a CP map $\Lambda^{(k)}$, taking operators acting on $\mathcal{H}_\rA \otimes \mathcal{H}_\rB$ to operators acting on $\mathcal{H}_\mathrm{O} = \mathbb{C}^{d_k}$, where $d_k = |\mathcal{T}_k|$, with Kraus operators $A_n = \sum_i \ket{i}\bra{n}T_i$, where $\{\ket{i}\}$ forms a basis for $\mathcal{H}_\mathrm{O}$, and $\{\ket{n}\}$ a basis for $\mathcal{H}_\rA \otimes \mathcal{H}_\rB$. The output of this CP map, when the input is a state $\rho^{\rA\rB}$ is
\begin{equation}\label{NPA channel moments}
\Gamma^{(k)} = \Lambda^{(k)}(\rho^{\rA\rB}) = \sum_{i,j}\ket{i}\bra{j} \tr[(T_j^\dagger T_i) \rho^{\rA\rB}].
\end{equation}
Since  $\Lambda^{(k)}$ preserves positivity, the output $\Gamma^{(k)}$ is positive-semidefinite. This output is referred to as a \emph{moment matrix}, for reasons that will become clearer below.  

By the nature of its construction, there are three types of conditions satisfied by the moment matrix:
\begin{enumerate}[(i)]
\item Certain elements of the moment matrix are accessible in a steering test, as they involve the moments $\expect{A_x^m B_y^n}$ \eqref{e:moments}. More specifically, whenever $T_j^\dagger T_i$ (or a linear combination thereof) can be expressed as a linear combination of  moments for each Alice and Bob,
\begin{multline}
\sum_{ij}\xi_{ij} T_j^\dagger T_i = \alpha (\openone\otimes\openone) + \sum_{x,m}\beta_{xm} (A_{x}^m\otimes \openone) \\ 
+ \sum_{y,n}\gamma_{yn} (\openone \otimes B_{y}^n) + \sum_{xy,mn} \delta_{xymn}(A_{x}^n\otimes B_{y}^n),
\end{multline}
for some coefficients $\{\xi_{ij}, \alpha, \beta_{xm}, \gamma_{yn}, \delta_{xymn}\}$, then, by defining $\Xi := \sum_{ij} \xi_{ij}\ket{j}\bra{i}$
\begin{equation}
\tr[\Xi\Gamma^{(k)}] = \Theta(\expect{A_x^m B_y^n}),
\end{equation}
where
\begin{multline}\label{def:Theta}
\Theta(\expect{A_x^m B_y^n}) := \alpha + \sum_{x,m}\beta_{xm}\expect{A_x^m} \\+ \sum_{y,n}\gamma_{yn}\expect{B_y^n} + \sum_{xy,mn}\delta_{xymn}\expect{A_x^m B_y^n}
\end{multline}
and where we have used the shorthand $\expect{A_x^m} \equiv \expect{A_x^m\otimes\openone}$ and $\expect{B_y^n} \equiv \expect{\openone \otimes B_y^n}$. 

\item Similarly to above, certain parts of the moment matrix contain only operators on Bob, and therefore are also known in a steering scenario. That is, if a linear combination of $T_j^\dagger T_i$ can be expressed only in terms of operators on Bob,
\begin{align}
\sum_{ij}\omega_{ij} T_j^\dagger T_i = \mathbb{P}(B_0,\ldots, B_{m_\rB}),
\end{align}
where $\mathbb{P}(B_0,\ldots, B_{m_\rB})$ is an arbitrary polynomial in the (non-commuting) observables $B_y$, then, defining $\Omega := \sum_{ij}\omega_{ij} \ket{j}\bra{i}$,
\begin{equation}
	\tr[\Omega\Gamma^{(k)}] = \expect{\mathbb{P}(B_0,\ldots, B_{m_\rB})}.
\end{equation} 
This encodes the fact that by assumption Bob can estimate the expectation of any operator $B$ on his system, in this case with $B = \mathbb{P}(B_0,\ldots, B_{m_\rB})$. 

\item Finally, since the (known) observables of Bob generate an algebra, \ie have equalities amongst products of themselves, this additionally places constraints among elements of the moment matrix. More precisely, whenever a relation of the form
\begin{equation}
\sum_{ij} \upsilon_{ij}T_j^\dagger T_i = 0,
\end{equation}
follows from the algebra of Bob, then, with $\Upsilon:= \sum_{ij}\upsilon_{ij}\ket{j}\bra{i}$
\begin{equation}
\tr[\Upsilon\Gamma^{(k)}] = 0.
\end{equation}
Note, in particular, that such relations can exist even between the elements of the moment matrix which are not directly observable in a steering test.
\end{enumerate}
The above three conditions must be satisfied by any moment matrix which is consistent with the data available in a (CV) steering test. That is, the above conditions are analogous to the conditions \eqref{nosignaling} and \eqref{normalization} satisfied by all valid assemblages. To create a test for steering, additional structure that is also necessarily satisfied when Alice and Bob share a separable state (and hence cannot demonstrate steering) needs to be added, \ie the analogue of Eq.~\eqref{LHS}. 

The final constraint can be added, using the fact, as outlined in Sec.~\ref{Subsec: SDI ent} that steering can only be demonstrated if the observables of Alice are incompatible (as well as the state being entangled). In the present context, this is equivalent to requiring that the observables of Alice necessarily do not all pairwise commute. To see this is equivalent to the more general notion of being non-jointly measurable given in Sec.~\ref{Subsec: SDI ent}, notice that the LHS assemblage $\sigma_{a|x} = \sum_\lambda D(a|x,\lambda)\sigma_\lambda$ can be reproduced by the state $\rho^{\rA\rB} = \sum_\lambda \ket{\lambda}\bra{\lambda}\otimes \sigma_\lambda$ and measurements $M_{a|x} = \sum_{\lambda'} D(a|x,\lambda')\ket{\lambda'}\bra{\lambda'}$, such that the associated observables $A_x = \sum_a aD(a|x,\lambda')\ket{\lambda'}\bra{\lambda'}$ commute, $[A_x,A_{x'}] = 0$. Thus, to generate a steering test, a final constraint can be imposed on the moment matrix
\begin{enumerate}[(i)]\setcounter{enumi}{3}
\item Whenever $T_j^\dagger T_i - T_{j'}^\dagger T_{i'} = 0$ due to $[A_x,A_x'] = 0$, then with $\Delta:= \ket{j}\bra{i} - \ket{j'}\bra{i'}$
\begin{equation}\label{e:commuting}
\tr[\Delta \Gamma^{(k)}] = 0
\end{equation}
In particular, this says that the moment matrix must be consistent with all the observables of Alice being mutually commuting. 
\end{enumerate}
Thus, the following is a test for steering: Given the data available in a steering test, determine whether there exists a valid moment matrix (of order $k$) consistent with it, satisfying the properties (i) -- (iv). Only if such a moment matrix exists could the data have come from measurements on a separable state. If on the other hand no such moment matrix exists, this certifies that steering has been witnessed -- i.e. that Alice measured non-commuting observables on an entangled state. More formally, the steering test is \cite{KSC+15}:
\begin{align}\label{e:CV SDP feas}
\text{given}&\quad \{\expect{A_x^m B_y^n}\}_{xymn}, \{\expect{\mathbb{P}_\ell(B_1,\ldots,B_{m_\rB})}\}_\ell, k \nonumber \\
\text{find}&\quad \Gamma^{(k)} \\
\text{s.t.}&\quad \tr[\Xi_i\Gamma^{(k)}] = \Theta_i(\expect{A_x^m B_y^n})\quad \forall i, \nonumber \\
&\quad \tr[\Omega_\ell\Gamma^{(k)}] = \expect{\mathbb{P}_\ell(B_0,\ldots, B_{m_\rB})}\quad \forall \ell, \nonumber \\
& \quad \tr[\Upsilon_j\Gamma^{(k)}] = 0 \quad \forall j,\quad \tr[\Delta_l\Gamma^{(k)}] = 0\quad \forall l,\nonumber \\
& \quad\Gamma^{(k)}\geq 0, \nonumber
\end{align}
where the first four sets of constraints are implicitly understood to hold for all independent sets (labelled by the corresponding roman character). 

Finally, from the duality theory of SDP, the above test generates steering inequalities in the CV setting, which depend only upon the observed data in the steering scenario, $\{\expect{A_x^m B_y^n}\}_{xymn}$, and $\{\expect{\mathbb{P}_\ell(B_1,\ldots,B_{m_\rB})}\}_\ell$. In particular, using the theory presented in Appendix \ref{SDP}, the dual of \eqref{e:CV SDP feas} is seen to be
\begin{align}
\text{given}&\quad \{\expect{A_x^m B_y^n}\}_{xymn}, \{\expect{\mathbb{P}_\ell(B_1,\ldots,B_{m_\rB})}\}_\ell, k \nonumber \\
\min_{\substack{w_i,x_\ell \\ y_j, z_l}}&\quad \sum_i w_i \Theta_i(\expect{A_x^m B_y^n}) + \sum_\ell x_\ell \expect{\mathbb{P}_\ell(B_0,\ldots, B_{m_\rB})} \nonumber \\
\text{s.t.}&\quad \sum_i w_i \Theta_i(\expect{A_x^m B_y^n}) + \sum_\ell x_\ell \expect{\mathbb{P}_\ell(B_0,\ldots, B_{m_\rB})} \nonumber \\
&\quad\quad +\sum_j y_j \Upsilon_j + \sum_l z_l\Delta_l \geq 0.
\end{align}
That is, $\{w_i\}_i$ and $\{x_\ell\}_\ell$ are the coefficients of the steering inequality formed out of $\Theta_i(\expect{A_x^m B_y^n})$ and $\expect{\mathbb{P}_\ell(B_0,\ldots, B_{m_\rB})}$. Since $\Theta_i(\cdot)$ is a linear function (see Eq.~\ref{def:Theta}), the steering inequalities obtained are linear in the joint expectation values, but depend non-linearly on the observables of Bob's system, through $\{\expect{\mathbb{P}_\ell(B_1,\ldots,B_{m_\rB})}\}_\ell$.

This approach was used in \cite{KSC+15} to demonstrate the steering of CV entangled states. In particular, it was shown that the lossy single photon state, $\rho = \eta\ket{\psi_1}\bra{\psi_1} + (1-\eta)\ket{00}\bra{00}$, where $\ket{\psi_1} = (\ket{01} + \ket{10})/\sqrt{2}$ is the single-photon NOON state (written in the Fock basis), and $\ket{00}$ is the two-mode vacuum, demonstrates steering if Alice and Bob perform homodyne measurements (and Bob estimates his local state), as long as $\eta > 2/3$, using the second level test $\Gamma^{(2)}$.  

To make the method reviewed above more concrete, we now present an example which should make all the abstract definitions clear.

\subsubsection{Example: Quadrature measurements}
Let us consider, as a concrete example, the case where Alice and Bob both measure two observables, and where the observables of Bob are $B_0 = \hat{q}$, and $B_1 = \hat{p}$, the position and momentum observables, \ie that Bob performs quadrature measurements. Let us also consider the case $k = 1$, such that
\begin{equation}
\mathcal{T}_1 = \{ \openone \otimes \openone, A_0 \otimes \openone, A_1 \otimes \openone, \openone \otimes \hat{q}, \openone \otimes \hat{p}\}
\end{equation}
and
\begin{equation}\label{e:example Gamma}
\Gamma^{(1)} =
\begin{pmatrix}
1 & \expect{A_0} & \expect{A_1} &  \expect{\hat{q}} & \expect{\hat{p}} \\
\expect{A_0} & \expect{A_0^2} & \expect{A_1A_0} &  \expect{A_0\hat{q}} & \expect{A_0\hat{p}} \\
\expect{A_1} & \expect{A_0A_1} & \expect{A_1^2} &  \expect{A_1\hat{q}} & \expect{A_1\hat{p}} \\
\expect{\hat{q}} & \expect{A_0\hat{q}} & \expect{A_1\hat{q}} &  \expect{\hat{q}^2} & \expect{\hat{p}\hat{q}} \\
\expect{\hat{p}} & \expect{A_0\hat{p}} & \expect{A_1\hat{p}} &  \expect{\hat{q}\hat{p}} & \expect{\hat{p}^2} \\
\end{pmatrix}.
\end{equation}
Notice that, apart from $\expect{A_0A_1}$, $\expect{A_1A_0}$, $\expect{\hat{q}\hat{p}}$ and $\expect{\hat{p}\hat{q}}$, all elements of $\Gamma^{(1)}$ are moments of the form \eqref{e:moments}. These elements are thus accessible in a steering experiment, and are fixed (condition (i)). Furthermore, $\expect{\hat{q}\hat{p}} = \expect{\mathbb{P}(\hat{q},\hat{p})}$ (and similarly for $\expect{\hat{p}\hat{q}}$), and hence these elements are also accessible in a steering experiment\footnote{Here, for example, Bob could measure $\Re(\expect{\hat{p}\hat{q}})$ and $\Im(\expect{\hat{p}\hat{q}})$, to reconstruct $\expect{\hat{p}\hat{q}}$} (condition (ii)). Finally, by assumption of no steering we have $[A_0,A_1] = 0$, hence $\expect{A_0A_1} = \expect{A_1A_0}$ (condition (iv)). Thus, in total there is one unknown element in \eqref{e:example Gamma}, $\expect{A_0A_1}$, and the test for steerability in this case is given by
\begin{align}
\text{given}&\quad \expect{A_0},\expect{A_1},\expect{\hat{q}},\expect{\hat{p}},\expect{A_0^2},\expect{A_1^2}, \expect{\hat{q}^2},\expect{\hat{p}^2},\nonumber \\
&\quad \expect{A_0\hat{q}},\expect{A_0\hat{p}},\expect{A_1\hat{q}},\expect{A_1\hat{p}},\expect{\hat{q}\hat{p}},\expect{\hat{p}\hat{q}} \nonumber \\
\text{find}&\quad z \\
\text{s.t.}&\quad \begin{pmatrix}
1 & \expect{A_0} & \expect{A_1} &  \expect{\hat{q}} & \expect{\hat{p}} \\
\expect{A_0} & \expect{A_0^2} & z &  \expect{A_0\hat{q}} & \expect{A_0\hat{p}} \\
\expect{A_1} & z & \expect{A_1^2} &  \expect{A_1\hat{q}} & \expect{A_1\hat{p}} \\
\expect{\hat{q}} & \expect{A_0\hat{q}} & \expect{A_1\hat{q}} &  \expect{\hat{q}^2} & \expect{\hat{p}\hat{q}} \\
\expect{\hat{p}} & \expect{A_0\hat{p}} & \expect{A_1\hat{p}} &  \expect{\hat{q}\hat{p}} & \expect{\hat{p}^2} \\
\end{pmatrix} \geq 0. \nonumber
\end{align}
If a $z$ can be found which makes this partially-specified moment matrix positive-semidefinite, then $z = \expect{A_0A_1} = \expect{A_1A_0}$, and the full moment matrix is consistent with the data that does not demonstrate steering. If on the other hand there exists no $z$ such that the above moment matrix is positive semidefinite, (\ie if the above feasibility SDP is infeasible) then the observed data demonstrates steering. 

In \cite{KSC+15} it was shown that if the above test is applied to Gaussian CV states, written in standard form, then it coincides with the the criteria previously obtained by Wiseman \etal \cite{WJD07} for the steerability of Gaussian CV states by Gaussian measurements.

\subsection{Further results on continuous-variable quantum steering}\label{cv inequalities}

The study of CV steering dates back to 1989 \cite{R89}. This work proposed an inequality based on an inferred Heisenberg uncertainty relation for quadrature measurements which was later recognised to be a steering inequality \cite{CJWR09}. The intuition behind this approach is that Bob can observe a violation of an uncertainty relation by conditioning on Alice's measurement results, which can only happen if they share entanglement. Thus, in principle one can derive steering inequalities for the discrete and continuous-variable cases based on any known uncertainty relation \cite{CR07,CJWR09}. Building upon this intuition the authors of Ref.~\cite{WRS+12} developed a steering inequality based on second moments, which in turned was used in the first experimental demonstration of steering free of loopholes. Similarly, entropic uncertainty principles were also used for steering detection \cite{WSG+11,SBW+13,ZHC15}.

Refs. \cite{R13,AS16} studied how bipartite steering can be demonstrated in a multipartite system, and showed monogamy constraints in the case of Gaussian states. More general constructions of continuous-variable steering inequalities were developed in \cite{JW11}. Finally, we will briefly review in Sec.~\ref{experiments} some experimental tests of steering in the continuous-variable regime.

\section{Applications}
\label{Sec: applications}

As we have seen before, the certification of steering implies the existence of a shared entangled state between Alice and Bob and the implementation of incompatible measurements in Alice's side. Intuitively, these properties imply other features in the setup, for example some degree of unpredictability of Alice's outcomes, or the fact that there is no third party that shares a maximally entangled state with the systems of Alice or Bob (this is often called the monogamy of entanglement). Because of this, the observation of an assemblage that demonstrates steering can be used to bound the usefulness of the setup in information theoretic tasks in the one-sided device independent scenario. In this section we review some of these applications whose figures of merit use SDPs based on steering.

{\subsection{One-sided device-independent entanglement estimation}{
As discussed in Sec.~\ref{Subsec: SDI ent}, the observation of an assemblage that demonstrates steering certifies that the underlying state shared by Alice and Bob used to produce it is entangled. It turns out that the assemblage also provides quantitative information about the amount of entanglement, as described below.

One way to quantify the amount of entanglement in a given state is through robustness-based quantifiers. These quantify entanglement through the minimal amount of certain types of noise needed to be added to the state before it becomes separable. In particular, robustness based quantifiers have the form
\begin{align} \label{eq: ent robustness}
\ER^{\mathcal{N}_e}(\rho^{\rA\rB}) = \min_{\sigma^{\rA\rB}, \rho^{\rA\rB}_{\sep}, r} & \quad r \\
\text{s.t.}& \quad \frac{\rho^{\rA\rB} + r \sigma^{\rA\rB}}{1+r} = \rho_\sep^{\rA\rB}, \nonumber\\
	& \quad \rho^{\rA\rB}_\sep \in \SEP, \quad \sigma^{\rA\rB} \in \mathcal{N}_e, \nonumber
\end{align}
where $\mathcal{N}_e$ is a convex set of states, which specifies the type of noise used to define the robustness. Common examples include $\mathcal{N}_e = \openone^{\rA\rB}/d_\rA d_\rB$ (Random Robustness), $\mathcal{N}_e = \SEP$ (Robustness) and $\mathcal{N}_e = \mathcal{S}$, the set of all states (Generalised Robustness). 

One can naturally make an association between given robustness based quantifiers of entanglement and steering. In particular, given the set $\mathcal{N}_e$ of states that characterises the robustness quantifier, one can find the corresponding set of  assemblages $\mathcal{N}$, those that can arise from measurements on states in $\mathcal{N}_e$: $\mathcal{N} = \{\sigma_{a|x} | \sigma_{a|x} = \tr_\rA[(M_{a|x} \otimes \openone^\rB)\sigma^{\rA\rB}],\sigma^{\rA\rB} \in \mathcal{N}_e, M_{a|x} \geq 0, \sum_a M_{a|x} = \openone\}$. 

Now, it is a straightforward observation that the steering robustness (with respect to the set $\mathcal{N}$) of any assemblage $\sigma_{a|x} = \tr_\rA[(M_{a|x} \otimes \openone^\rB)\rho^{\rA\rB}]$ that arises from measurements $M_{a|x}$ on a state $\rho^{\rA\rB}$ provides a one-sided device-independent lower bound on the entanglement robustness (with respect to $\mathcal{N}_e$). Indeed, consider that $\sigma^{\rA\rB*}$ is the state which attains the minimum $r^*$ in \eqref{eq: ent robustness}. Then, by definition $\pi_{a|x} = \tr_\rA[(M_{a|x} \otimes \openone^\rB)\sigma^{\rA\rB*}] \in \mathcal{N}$, and hence
\begin{equation}
	\frac{\sigma_{a|x} + r^* \pi_{a|x}}{1+r^*} = \sigma_{a|x}^\LHS.
\end{equation}
Thus it follows that $\SR^\mathcal{N}(\sigma_{a|x}) \leq r^* = \ER^{\mathcal{N}_e}(\rho^{\rA\rB})$, which establishes the lower bound. Clearly, by optimising over the set of measurements $M_{a|x}$ one could search for the best possible one-sided device-independent lower bound on the entanglement robustness. We also note that exactly the same argument works for the Best-Separable-Approximation \cite{LS98} entanglement quantifier, for which the steering weight \eqref{eq: SW} provides a one-sided device-independent lower bound.  

One can also go beyond the robustness based quantifiers, although in a less direct manner, by using the bipartite version of the steering NPA hierarchy of moment matrices \cite{P13}, discussed in Sec.~\ref{quantumness testing}. In particular, consider now the set $\mathcal{M} = \{M_{a|x} \}_{a,x}$ of all measurement operators for Alice, (which we recall can without loss of generality be taken to be projective). Starting from this set, the set of all strings of operators of length $k$ or less can be formed,
\begin{align}
\mathcal{S}_0 &= \{\openone\}, \nonumber \\
\mathcal{S}_k &= \mathcal{S}_{k-1} \cup \{M_i S_j | S_j \in \mathcal{S}_{k-1}, M_i \in \mathcal{M}\}.
\end{align}
In general, this set of operators will satisfy certain relations of the form
\begin{equation}\label{e:NPA consist Pusey}
\begin{split}
\sum_{ij}\xi_{ij} S_j^\dagger S_i &= \alpha \openone + \sum_{ax}\beta_{ax} M_{a|x}, \\
\sum_{ij}\upsilon_{ij} S_j^\dagger S_i &= 0,
\end{split}
\end{equation}
for some coefficients $\{\xi_{ij}, \alpha, \beta_{ax}\}$ and $\{ \upsilon_{ij}\}$. That is, some combination of operators will depend upon the measurement operators, whilst other combinations will identically vanish.  

As previously, we consider the operator $\Gamma^{(k)}$, which is the output of a CP map $\Lambda^{(k)}\otimes \mathrm{id}(\cdot)$, taking $\mathcal{H}_\rA \otimes \mathcal{H}_\rB$ to $\mathcal{H}_\rO \otimes \mathcal{H}_\rB$, where $\Lambda^{(k)}(\cdot) = \sum_k A_k(\cdot)A_k^\dagger$ and $A_k = \sum_i \ket{i}\bra{k}S_i$. The output contains the members of the assemblage $\sigma_{a|x}$ as blocks, and moreover, by virtue of \eqref{e:NPA consist Pusey} satisfies the constraints
\begin{equation}\label{e:NPA relations}
\begin{split}
	\tr_\rO[(\Xi \otimes \openone_\rB)\Gamma^{(k)}] &= \Theta(\sigma_{a|x}), \\
	\tr_\rO[(\Upsilon \otimes \openone_\rB)\Gamma^{(k)}] &= 0, 
	\end{split}
\end{equation}
where $\Xi = \sum_{ij}\xi_{ij}\ket{j}\bra{i}$, $\Upsilon = \sum_{ij}\upsilon_{ij}\ket{j}\bra{i}$ and $\Theta(\sigma_{a|x}) = \alpha \rho^{\rB} + \sum_{a,x} \beta_{ax}\sigma_{a|x}$. The crucial point is that for any valid observable data from a steering test, it must be possible to find a completion of $\Gamma^{(k)}$ that is positive semidefinite. By itself this is not yet useful, since from the GHJW theorem \cite{G89,HJW93} we already know that all valid non-signalling assemblages have a quantum realisation, and hence every operator $\Gamma^{(k)}$ has a completion. 

However, the moment matrix becomes useful when we consider subsets of quantum states. In particular, one would like to characterise those assemblages that can arise from measurements on a PPT state, \ie those that satisfy $(\rho^{\rA\rB})^{\rT_\rB} \geq 0$. Then, it follows that the output of the CP map $\Lambda^{(k)}\otimes \mathrm{id} (\rho^{\rA\rB})$ is also PPT, for all $\rho^{\rA\rB}$ PPT. That is, the $\Gamma^{(k)}$ which can arise from PPT states pick up additional structure relative to the most general $\Gamma^{(k)}$. This thus provides a necessary condition for a given assemblage to be compatible with having arisen from a PPT state \cite{P13}. 

As an application of the above one can get one-sided device-independent lower bounds on the negativity of a quantum state, $N(\rho^{\rA\rB}) = (\left\| (\rho^{\rA\rB})^{\rT_\rB}\right\|_1 - 1)/2$ \cite{VW02}. The negativity can alternatively be written as an SDP:
\begin{align}
	N(\rho^{\rA\rB}) = \min_{\sigma_+,\sigma_-} & \quad \tr[\sigma_-] \\
\text{s.t.}& \quad \rho^{\rA\rB} = \sigma_+ - \sigma_- \nonumber \\
& \quad \sigma_+^{\rT_\rB} \geq 0,\quad \sigma_-^{\rT_\rB} \geq 0. \nonumber
\end{align}
To obtain a lower bound, we note first, given that $S_0 = \openone$, it follows that $(\bra{0}\otimes \openone )\Gamma^{(k)}(\ket{0}\otimes \openone) = \rho^{\rB}$ and hence $\tr[(\ketbra{0}{0}\otimes \openone)\Gamma^{(k)}] = \tr[\rho^\rB]$. Thus, by associating $\Gamma^{(k)}_{\pm}$ with $\sigma_{\pm}$, we find that
\begin{align}
	N(\rho^{\rA\rB}) \leq \min_{\Gamma_+^{(k)}, \Gamma_-^{(k)}} & \quad \tr[(\ketbra{0}{0}\otimes \openone)\Gamma^{(k)}_-] \\
\text{s.t.}& \quad \Gamma^{(k)} = \Gamma^{(k)}_+ - \Gamma^{(k)}_-, \nonumber \\
& \quad \left(\Gamma^{(k)}_+\right)^{\rT_\rB} \geq 0,\quad \left(\Gamma^{(k)}_-\right)^{\rT_\rB} \geq 0, \nonumber \\
& \quad \tr_\rO[(\Xi_i \otimes \openone_\rB)\Gamma^{(k)}] = \Theta_i(\sigma_{a|x}), \nonumber \\
& \quad	\tr_\rO[(\Upsilon_i \otimes \openone_\rB)\Gamma^{(k)}] = 0, \nonumber \\
& \quad	\tr_\rO[(\Upsilon_i \otimes \openone_\rB)\Gamma^{(k)}_\pm] = 0, \nonumber
\end{align}
where the last three constraints are understood to hold for all independent relations of the form \eqref{e:NPA relations}. The above technique was introduced in Ref. \cite{P13} as a method to study the `stronger Peres conjecture', whether or not PPT entangled states can demonstrate steering or not\footnote{The Peres conjecture stated the PPT entangled states should be incapable of demonstrating nonlocality. This has recently been proven to be false, by explicit counter-example \cite{VB14}.}. Although Ref.~\cite{P13} did not manage to come to definite conclusions, subsequent work (using a different technique), managed to show that PPT entangled states can indeed demonstrate steering \cite{MGHG14}.

In Ref.~\cite{TMG15} it was shown furthermore how one can lower bound in a one-sided device independent manner any entanglement measure which is defined through a convex-roof extension. The authors again use the moment matrix to map from quantum states, to the partial information that is accessible in a steering experiment. The key additional requirement is that the CP map $\Lambda(\cdot)$ needs to be constructed to not only be completely-positive, but moreover trace-preserving, since local trace-preserving completely-positive maps cannot increase entanglement monotones. Trace-preservation can be achieved by taking the set
\begin{equation}\label{e:S tp}
	\mathcal{S} = \{M_{a_0|x=0} M_{a_1|x=1} \cdots M_{a_{m_\rA-1}|x=m_\rA}-1\}_{a_0,a_1,\ldots,a_{m_\rA-1}},
\end{equation} 
which still has structure of the form \eqref{e:NPA consist Pusey} and hence will lead to relations of the form \eqref{e:NPA relations} on the CP map $\Lambda(\cdot)$.

The main idea is to show how entanglement measures based upon convex-roof extensions can be evaluated as a expectation value of an operator on a multipartite separable state. As an example, consider the linear entropy of entanglement, which for pure states is $E_\mathrm{lin}\big(\ket{\psi}^{\rA\rB}\big) = S_\mathrm{lin}(\tr_\rA[\ketbra{\psi}{\psi}^{\rA\rB}])$, with $S_\mathrm{lin}(\rho) = 1 - \tr[\rho^2]$ the linear entropy. For mixed states it is given by the convex-roof
\begin{equation}
E_\mathrm{lin}(\rho^{\rA\rB}) = \min_{\{p_k, \ket{\psi_k}^{\rA\rB}\}} \sum_k p_k E_\mathrm{lin}\big(\ket{\psi_k}^{\rA\rB}\big),
\end{equation}
where the minimisation is over all pure-state decompositions $\rho^{\rA\rB} = \sum_k p_k \ketbra{\psi_k}{\psi_k}^{\rA\rB}$. In Ref.~\cite{TMG15} it is shown that if $\{\tilde{p}_k, \ket{\tilde{\psi}_k}\}$ is the decomposition achieving the minimum, then the linear entropy can alternatively be written
\begin{equation}
	E_\mathrm{lin}(\rho^{\rA\rB}) = \tr[(\mathcal{A}^{\rA\rA'}\otimes \openone^{\rB\rB'})\omega^{\rA\rA'\rB\rB'}]
\end{equation}
where $\omega^{\rA\rA'\rB\rB'} = \sum_k \tilde{p}_k \ketbra{\tilde{\psi}_k}{\tilde{\psi}_k}^{\rA\rB} \otimes \ketbra{\tilde{\psi}_k}{\tilde{\psi}_k}^{\rA'\rB'}$, and $\mathcal{A}^{\rA\rA'}$ is the projector onto the antisymmetric subspace of $\mathcal{H}_\rA \otimes \mathcal{H}_\rA'$. Crucially, $\omega^{\rA\rA'\rB\rB'}$ is separable across the bipartition $\rA\rB$:$\rA'\rB'$, and satisfies $\tr_{\rA\rB}[\omega^{\rA\rA'\rB\rB'}] = \tr_{\rA'\rB'}[\omega^{\rA\rA'\rB\rB'}] = \rho^{\rA\rB}$, \ie it is a symmetric extension of the state \cite{DPS02}. Put together, this shows that the linear entropy of entanglement of a state can be cast as the following optimisation problem
\begin{align}
	E_\mathrm{lin}(\rho^{\rA\rB}) = \min_{\omega^{\rA\rA'\rB\rB'}}&\quad \tr[(\mathcal{A}^{\rA\rA'}\otimes \openone^{\rB\rB'})\omega^{\rA\rA'\rB\rB'}] \\
	\text{s.t.}&\quad \omega^{\rA\rA'\rB\rB'} \in \mathrm{SEP}_{\rA\rB:\rA'\rB'},\nonumber \\
	&\quad \tr_{\rA\rB}[\omega^{\rA\rA'\rB\rB'}] = \rho^{\rA\rB}, \nonumber \\
	& \quad \tr_{\rA'\rB'}[\omega^{\rA\rA'\rB\rB'}] = \rho^{\rA\rB}. \nonumber
\end{align}
Although this problem is not natively an SDP, similarly to Appendix.~\ref{Separability testing}, it can easily be relaxed to an SDP, by replacing the constraint on separability to positivity under partial transposition, or to $k$-extendibility. This thus provides an SDP lower bound on the linear entropy of entanglement. To move to the one-sided device-independent scenario, one replaces $\rho^{\rA\rB}$ by $\Gamma = \Lambda_\rA \otimes \mathrm{id}_\rB (\rho^{\rA\rB})$, which is partially known in a steering scenario. Here the set of operators $\mathcal{S}$ from \eqref{e:S tp} is used, which ensure that $\Lambda_\rA(\cdot)$ is trace-preserving, Since trace-preserving, local completely positive maps cannot increase entanglement, the entanglement of $\Gamma$ cannot be more than that of $\rho^{\rA\rB}$. We thus obtain the one-sided device-independent lower bound on the linear entropy of entanglement
\begin{align}
	E_\mathrm{lin}(\rho^{\rA\rB}) \leq \min_{\omega^{\rA\rA'\rB\rB'},\Gamma}&\quad \tr[(\mathcal{A}^{\rA\rA'}\otimes \openone^{\rB\rB'})\omega^{\rA\rA'\rB\rB'}] \nonumber \\
	\text{s.t.}&\quad \omega^{\rA\rA'\rB\rB'} \in \mathrm{SEP}_{\rA\rB:\rA'\rB'}, \\
	&\quad \tr_{\rA\rB}[\omega^{\rA\rA'\rB\rB'}] = \Gamma, \nonumber \\
	& \quad \tr_{\rA'\rB'}[\omega^{\rA\rA'\rB\rB'}] = \Gamma, \nonumber \\
	& \quad \tr_\rO[(\Xi_i \otimes \openone_\rB)\Gamma^{(k)}] = \Theta_i(\sigma_{a|x}), \nonumber \\
& \quad	\tr_\rO[(\Upsilon_i \otimes \openone_\rB)\Gamma^{(k)}] = 0, \nonumber
\end{align}
which again can be itself lower bounded by an SDP by replacing the separability constraint, as above. Ref.~\cite{TMG15} furthermore shows how other convex-roof entanglement measures can similarly be bounded in a one-sided device-independent manner, and show how in certain circumstances a slightly improved method can be employed to obtain better bounds, the details of which are bound the scope of the present review. }

\subsection{One-sided device-independent estimation of measurement incompatibility}\label{Subsec: measurement incompatibility}

The observation of an assemblage that demonstrates steering also certifies that the measurements that Alice used to produce it are incompatible, \ie can not be performed jointly \cite{BLM96}, as discussed in Sec.~\ref{Sec: definition}. Again, it turns out that the assemblage also provides quantitative information about the amount of measurement incompatibility in Alice's measurements \cite{CS16} as explained below.

Similarly to the quantifiers of steering described before, one can also define robustness-based quantifiers of measurement incompatibility as the minimal amount of certain types of noise needed to be added to a set of measurements such that they become jointly measurable \cite{HKRS15,P15,UBGP15}. In particular one can define the following incompatibility quantifiers:

\begin{itemize}
\item The \textit{Incompatibility Weight} of a set measurements $M_{a|x}$ is based on a decomposition of the measurements $M_{a|x}$ in terms of a convex combination of an arbitrary set of measurement $O_{a|x}$ and an arbitrary set of jointly-measurable measurements $N_{a|x}$ \cite{P15}. It thus quantify the maximal weight of the jointly measurable set $N_{a|x}$ that can be used in such a decomposition.
\item The \textit{Incompatibility Robustness} of a set of measurements $M_{a|x}$ is the minimal $t$ such that there exist another set of measurements $N_{a|x}$ for which the mixture $(M_{a|x}+t N_{a|x})/(1+t)$ defines a jointly measurable measurement \cite{UBGP15}.
\item The \textit{Incompatibility Random Robustness} \cite{HKRS15} is a particular case of the incompatibility robustness above where $N_{a|x}=\openone/{o_\rA}$, and $o_\rA$ is the number of outcomes of the measurements $M_{a|x}$.

\item The \textit{Incompatibility Jointly-Measurable Robustness} is another particular case of the incompatibility robustness above, where now $N_{a|x}$ must be a set of jointly-measurable measurements.
\end{itemize}

In Ref.~\cite{CS16} is was shown, by slightly modifying the steering quantifiers defined in Sec.~\ref{Sec: detecting}, that one arrives at new steering quantifiers that provide lower bounds to the above measurement incompatibility quantifiers. Thus, by calculating these new quantifiers one obtains a one-sided device-independent estimation of the measurements that Alice performed. The new steering quantifiers are the Consistent Steering Weight, the Consistent Steering Robustness, the Reduced-State Steering Robustness and the Consistent LHS Steering Robustness, which provide lower bounds to the Incompatibility Weight, Incompatibility Robustness, Incompatibility Random Robustness and Incompatibility Jointly Measurable Robustness respectively. They are defined as follows:

\begin{itemize}
\item The \textit{Consistent Steering Weight} of an assemblage $\sigma_{a|x}$ is defined as the maximal $p$ such that $\sigma_{a|x}=(1-p)\gamma_{a|x}+p\sigma_{a|x}^\LHS$, where $\sigma_{a|x}^\LHS$ is an LHS assemblage with the property that $\sum_{a} \sigma^\LHS_{a|x} = \sum_{a} \sigma_{a|x}\qu \forall x$. In other words, the assemblage $\sigma_{a|x}^\LHS$ (and consequently $\gamma_{a|x}$) defines the same reduced state for Bob's system as $\sigma_{a|x}$.
\item The \textit{Consistent Robustness} of the assemblage $\sigma_{a|x}$ is the minimal $t$ such that there exist another assemblage $\pi_{a|x}$ satisfying $\sum_{a} \sigma_{a|x} = \sum_{a} \pi_{a|x}\qu\forall x$, for which the mixture $(\sigma_{a|x}+t \pi_{a|x})/(1+t)$ defines an LHS assemblage. Again, this means that  $\pi_{a|x}$ and $\sigma_{a|x}$ define the same reduced state.
\item The \textit{Reduced-State Robustness} of the assemblage $\sigma_{a|x}$ is the minimal $t$ such that the assemblage $(\sigma_{a|x}+t \sum_a \sigma_{a|x})/(1+t)$ is LHS.

\item The \textit{Consistent LHS Robustness} of the assemblage $\sigma_{a|x}$ is the minimal $t$ such that there exist another LHS assemblage $\pi_{a|x}^\LHS$ satisfying $\sum_{a} \sigma_{a|x} = \sum_{a} \pi_{a|x}^\LHS\qu\forall x$, for which the mixture $(\sigma_{a|x}+t \pi_{a|x}^\LHS)/(1+t)$ defines an LHS assemblage. Again, this means that  $\pi_{a|x}$ and $\sigma_{a|x}$ define the same reduced state.

\end{itemize}

\subsection{Sub-channel discrimination}\label{Subsec: subchannel disc}

Another task which turns out to be closely related to steering is sub-channel discrimination \cite{PW15}. This is a protocol where Alice sends a state to Bob via a channel $\Lambda(\cdot)$, that is composed by several branches  $\Lambda_a(\cdot)$, such that $\Lambda(\rho)=\sum_a \Lambda_a(\rho)$. The task is to figure out which branch $\Lambda_a(\cdot)$ has acted on the state. The optimal success probability in this task depends on the resources that Alice and Bob are allowed to use. Piani and Watrous proved that if they are restricted to local measurements assisted by one-way communication, then for every steerable state $\rho^{\rA\rB}$ there exist a one-sided channel $\mathrm{id}_\rA\otimes\Lambda(\cdot)$ for which this state provides an advantage over purely classical strategies in the task. Moreover this advantage is exactly quantified by the Steering Robustness $\SR(\rho^{\rA\rB})$ of the state $\rho^{\rA\rB}$, as defined in \eqref{e:SQ state}.

Notice that although the Steering Robustness of a state is not obtained by an SDP (due to the maximisation over all assemblages obtainable from $\rho^{\rA\rB}$) one can perform a see-saw optimisation (see App. \ref{sec: see-saw}) in order to find better and better lower bounds on the robustness of $\rho^{\rA\rB}$. This works as follows. First one chooses a set of measurements $M_{a|x}$ that leads to an assemblage $\sigma_{a|x}$, from which the steering robustness is computed though \eqref{eq: steering robustness dual}. If this assemblage is steerable, this code will produce operators $F_{a|x}^*$ such that $\tr \sum_{ax} F^*_{a|x}\sigma_{a|x}$ is the steering robustness of the assemblage $\sigma_{a|x}$. Now, one can run the same SDP, fixing the operators $F_{a|x}=F_{a|x}^*$, rewriting $\sigma_{a|x}=\Tr_\rA[(M_{a|x}\otimes\openone) \rho^{\rA\rB}]$, and optimising over $\{M_{a|x}\}_{a,x}$. That is, one can search for the best measurements $\{M_{a|x}\}_{a,x}$ for which the operators $F^*_{a|x}$ can detect the maximal steering robustness of the generated assemblage. By iterating this procedure one finds better and better lower bounds on the steering robustness of the state $\rho^{\rA\rB}$. 

\subsection{One-side device-independent randomness certification}\label{Subsec: randomness}

One of the most distinctive features of quantum mechanics is that the outcomes of quantum measurements can be intrinsically random, in the sense that their unpredictability can not be attributed to any lack of ignorance about the physical setup. Such randomness is at the basis of quantum random number generators. However, these devices usually rely on the modelling of the physical devices that generate the random numbers (for example, a single photon being split in a beam splitter and subsequently measured by efficient detectors). 

Recent results showed that quantum mechanics also allow us to certify a stronger form of randomness which does not rely on the description of the measuring devices, \ie in a device-independent way. This is often called intrinsic randomness and is certified through the violation of Bell inequalities \cite{C09,PAM+10,MPA11,BCP+14}. It turns out that intrinsic randomness can also be certified (and quantified) in the steering scenario \cite{ZTBS14,PCSA15}. This form of randomness can be used as a primitive for one-sided device-independent quantum key distribution \cite{BCW+12}. In what follows we will discuss the results of Ref.~\cite{PCSA15}, which used SDP to calculate the number of random bits generated by Alice's measurements based on the assemblage observed by Bob.

\subsubsection{Local Randomness}

We will start by reviewing the task of local randomness certification, where the  interest is in how unpredictable the outcomes of a measurement of one of the parties are. The main figure of merit is the probability $P_{\guess}(x^*)$ that a third party, called Eve, can guess the outcome $a$ of a measurement $x^*$ performed by Alice. The physical situation consists in Alice, Bob, and Eve initially sharing a tripartite state $\rho^{\rA\rB\rE}$ which could be prepared by Eve. It is also assumed that Eve knows the form of the measurements Alice can make, \ie she knows $\{M_{a|x}\}_{a,x}$. {At each round of the experiment, Eve applies a measurement $M_e$ on her share of the system, with the aim that her outcome $e$ equals that of Alice's $a$, whenever Alice measures $x^*$, with the highest probability}\footnote{Notice that here we are not considering the most general attack in which Eve could hold all the copies of her systems, and later apply a global measurement on them to guess all outcomes simultaneously.} In other words Eve wants to maximise her guessing probability
{
\begin{align}\label{e:pguess}
P_{\guess}(x^*)&= P(e=a|x^*)=\sum_a P(a,e=a|x^*)\nonumber\\
&=\sum_a \tr[(M_{a|x^*}\otimes \openone \otimes M_{e=a}) \rho^{\rA\rB\rE}].
\end{align}
}
If  $P_{\guess}(x^*)=1$, this means that the outcomes of Alice and Eve are perfectly correlated whenever Alice chooses $x=x^*$. However, if $P_{\guess}(x^*)<1$ Eve can not predict Alice's outcome with certainty, so these outcomes have some intrinsic randomness.

Notice that the states left for Bob after Alice and Eve's measurements are given by $\sigma_{a|x}^e=\tr_{\rA\rE}[(M_{a|x}\otimes \openone \otimes M_{e=a}) \rho^{\rA\rB\rE}]$ .  However, he will not know which outcome Eve has obtained, hence he observes $\sigma_{a|x}^{\obs}= \sum_e \sigma_{a|x}^e$. With this in mind,  the guessing probability \eqref{e:pguess} can be written in terms of Bob's assemblage, which allows one to write down an SDP that certifies how much knowledge Eve could have about Alice's outcomes:
\begin{align}\label{e:local guess}
P_{\guess}(x^*)=\max_{\{\sigma_{a|x}^e\}}&\quad \tr\sum_e \sigma_{a=e|x^*}^e \\
\text{s.t.} &\quad \sum_e \sigma_{a|x}^e = \sigma_{a|x}^{\obs}  \quad\forall a,x, \nonumber \\
&\quad\sum_a \sigma_{a|x}^e = \sum_a \sigma_{a|x'}^e \quad\forall e,x,x', \nonumber \\
&\quad\sigma_{a|x}^e \geq 0 \quad\forall a,x,e.\nonumber
\end{align}
The first constraint states that the assemblage that Bob observes is an average over the assemblages prepared by the measurement of Eve. The second assures that there is no signalling from Alice to Eve and Bob (signalling from Eve is ruled out from the fact that she has only performs a single measurement). Finally, the last condition assures that the states prepared by Alice and Eve are valid (positive semidefinite). Note that normalisation ($\tr\sum_{ae}\sigma_{a|x}^e = 1$) is automatic, from the first constraint. This SDP has the interpretation of maximising the guessing probability \eqref{e:pguess} among all possible quantum realisations that provide the assemblage $\sigma_{a|x}^{\obs}$ to Alice and Bob.

Finally, by using the duality theory of SDP, the guessing probability can alternatively be re-expressed in terms of the violation of a steering inequality. In particular, the dual to \eqref{e:local guess} is:
\begin{align}
P_{\guess}(x^*)=\min_{\substack{\{F_{a|x}\}\\\{G_x^e\}}}&\quad \tr\sum_{ax} F_{a|x}\sigma_{a|x}^\obs \\
\text{s.t.} &\quad F_{a|x} - \delta_{a,e}\delta_{x,x^*}\openone \nonumber \\
&\quad\quad- G_x^e + \delta_{x,x^*}\sum_{x'}G_{x'}^e \geq 0\quad \forall a,e,x.\nonumber
\end{align}
The constraint can be understood by multiplying by an arbitrary assemblage $\assem{\pi}{a|x}{}$, and taking the sum and trace, to arrive at $\tr\sum_{ax}F_{a|x}\pi_{a|x} \geq \tr[\pi_{e|x^*}] \qu \forall e$. Since $p(a=e|x^*) = \tr[\pi_{e|x^*}]$, this shows that the constraint enforces the property that the value of the inequality upper bounds the determinism of the the outcomes of the measurement $x^*$.

\subsubsection{Global Randomness}

A second task is to quantify the unpredictability of the outcome of both Alice and Bob. The fact that Bob uses characterised measurement devices may seem to make it irrelevant to consider how much information Eve can obtain about his measurement. However, notice that in the task Eve could still prepare a state that maximises her knowledge over Bob's results.

The first proposal to certify global randomness in the steering scenario used a similar strategy to the fully device-independent case \cite{ZTBS14}. The idea was to assume that Bob also applies some specific measurements and to bound the predictability of the eavesdropper by the knowledge of the nonlocal behaviour observed by Alice and Bob, utilising the known algebraic relations among the measurements Bob applies. For instance, if Bob performs two measurements $B_1$ and $B_2$ corresponding to Pauli measurements $X$ and $Y$,  this implies that the statistics he observes has to agree with the fact that $B_1 B_2=-B_2 B_1$. Although this method already provides an enhancement compared to fully device-independent randomness certification (because of the extra assumptions), it does not use all the information available in the steering scenario.

In what follows we review a method to quantify how much Eve can learn about a specific pair of measurements for Alice and Bob, given an observed assemblage.
Consider that Bob uses a measurement with POVM elements $\{M_b\}_b$ to extract randomness, while Alice uses the measurement $x = x^*$. Eve now has to give a pair of guesses $(e,e')$ for the pair $(a,b)$. The probability of Eve guessing both outcomes is given by the solution to the following SDP
\begin{align} \label{e:global pguess}
P_{\guess}(x^*,M_b) = \max_{\{\sigma_{a|x}^{ee'}\}} &\qu \tr\sum_{e e'} M_{b=e'} \sigma_{a=e|x^*}^{e e'}  \nonumber \\
 \text{s.t.} &\qu \sum_{ee'} \sigma_{a|x}^{ee'} = \sigma_{a|x}^\obs\quad\forall a,x,  \\
&\qu \sum_a \sigma_{a|x}^{ee'} = \sum_a \sigma_{a|x'}^{ee'} \qu \forall x,x',e,e' \nonumber \\
 &\qu \sigma_{a|x}^{ee'} \geq 0 \quad \forall a,x,e,e'.\nonumber
\end{align}
Comparing to the local randomness case \eqref{e:pguess} this SDP maximises the probability that Eve's outcomes equal those of Alice and Bob, optimised over all possible assemblages that Eve and Alice could prepare for Bob, considering that he makes the known measurement $M_b$. Once again the consistency with the observed assemblage $\sigma_{a|x}^\mathrm{obs}$, no-signalling and positivity are required. Finally, the dual formulation of \eqref{e:global pguess} gives the alternative formulation
\begin{align}
P_{\guess}(x^*,M_b)=\min_{\substack{\{F_{a|x}\}\\\{G_x^{ee'}\}}}&\quad \tr\sum_{ax} F_{a|x}\sigma_{a|x}^\obs \\
\text{s.t.} &\quad F_{a|x} - \delta_{a,e}\delta_{x,x^*}M_{e'} - G_x^{ee'}\nonumber \\
&\quad\quad + \delta_{x,x^*}\sum_{x'}G_{x'}^{ee'} \geq 0\quad \forall a,e,e',x.\nonumber
\end{align}
Similarly to the case of the local guessing probability, the constraint can be understood as enforcing $\tr\sum_{ax}F_{a|x}\pi_{a|x} \geq \tr[M_{e'}\pi_{e|x^*}] \qu \forall e,e'$. Since $p(a=e,b=e'|x^*) = \tr[M_{e'}\pi_{e|x^*}]$, this shows that the constraint enforces the property that the value of the inequality upper bounds the determinism of the the joint outcomes of the measurement $x^*$ of Alice, and $\{M_b\}_b$ of Bob. 

\subsection{One-sided device-independent self-testing}\label{Subsec: self testing}

Self-testing is a task which is customarily considered in the fully device-independent scenario (\ie the Bell scenario), where the parties observe the violation of a Bell inequality. The basic observation behind self-testing is that if one obtains the maximal violation of the CHSH Bell inequality then one can certify both that the state measured was (up to local isometries) the maximally entangled state of two qubits, and that the measurements were Pauli spin measurements \cite{SW87,PR92,T93,MY03}. More generally, from a modern perspective, one would like to bound the distance between the state actually produced and some target state, and the measurements actually implemented and some target measurements, given only the observed violation of a Bell inequality (see, e.g.~ \cite{MS12,MYS12,RUV13,YVB+14,BP15}). 

Motivated by these results, in Refs.~\cite{GWK15,SH16} the authors proposed the study of self-testing in the one-sided device-independent scenario (\ie using quantum steering). In particular, in \cite{SH16} the authors provide a SDP which, based on the violation of a given steering inequality, can lower bound the fidelity between the measured state and a target maximally entangled state.

\subsection{Maximal violation of Bell inequalities}\label{Subsec: max viol}

{Let us consider a generic N-partite Bell inequality $S\equiv\sum_{\textbf{a}b\textbf{x}y}c_{\textbf{a}b\textbf{x}y} P(\textbf{a}b|\textbf{x}y)\leq \beta$, where $\textbf{a}\equiv a_1,a_2,...,a_{N-1}$ and $\textbf{x}\equiv x_1,x_2,...,x_{N-1}$ is a shorthand notation to express the outcomes and inputs of the first $N-1$ parties, $b$ and $y$ labels the outputs and inputs of the $N^{th}$ party Bob, $c_{\textbf{a}b\textbf{x}y}$ are coefficients defining the inequality and $\beta$ the bound for behaviours having a local hidden variable (LHV) description \cite{BCP+14}. In this scenario, given that the first $N-1$ parties apply their measurements, the last party is left with the assemblage $\sigma_{\textbf{a}|\textbf{x}}$, from which one can calculate the maximum value that the Bell functional $S$ can achieve via the following SDP:
\begin{align}\label{e:max viol}
\text{given}&\quad\sigma_{\textbf{a}|\textbf{x}}\\
\max_{\{M_{b|y}\}} &\quad\sum_{\textbf{a}b\textbf{x}y}c_{\textbf{a}b\textbf{x}y} P(\textbf{a}b|\textbf{x}y)\nonumber\\
\text{s.t.} &\quad P(ab|xy)=\tr(M_{b|y}\sigma_{\textbf{a}|\textbf{x}})  \nonumber \\
&\quad M_{b|y}\geq0\quad\forall b,y, \quad \sum_{b}M_{b|y}=\openone~\forall y.\nonumber
\end{align}
}
Notice that an analytical solution for this problem was given in Ref. \cite{TRL16} in the case of Bell inequalities where Bob performs dichotomic measurements on a two-dimensional state.

It is also possible to use the see-saw approach described in Appendix~\ref{sec: see-saw} to find bounds to the maximal value of $S$ that can be achieved by restricting Bob's system to a given dimension. In order to do this one can first runs the SDP \eqref{e:max viol} for a given initially chosen assemblage and obtains the optimal measurements $M_{b|y}^*$ for this particular assemblage. In the second step one can then fix the measurements $M_{b|y}^*$ and run a similar SDP that searches for the best assemblage $\sigma_{\textbf{a}|\textbf{x}}^*$ given these measurements. By iterating this procedure one can find better and better lower bounds on $S$.

\subsection{Further applications in the steering scenario}\label{Subsec: other applications}

The first application in the steering scenario discussed in the literature (apart from entanglement certification) was one-side device-independent quantum key distribution. As noted in \cite{BCW+12}, it was shown in Refs.~\cite{TR11,TLGR12} how to bound the secret key rate by a function that is based only on the uncertainty relations satisfied by Bob's measurements. Thus, since to estimate the value of this function no assumptions need to be made about Alice's implementation, this function can be seen as a steering witness \cite{BCW+12} (see also Ref.~\cite{WWR14} for analogous results in the continuous-variable case). 

{Another security task related to steering is quantum secret sharing. Such a connection was first qualitatively notice in Ref.~\cite{HR13} and later on formalised in Ref.~\cite{XKAH16} (see also \cite{KXHA16}).

Finally, steering has also been shown to be related to quantum teleportation. In Ref.~\cite{QRAR15} the authors proved that any Gaussian state that is useful for secure teleportation is necessarily two-way steerable (\ie, these states can be used to demonstrate steering when either Alice or Bob are the untrusted parties).}

\section{Practical aspects}
\label{Sec: experimental}

In the last section of this article we will describe how to use some of the techniques presented before in situations that might be relevant for experimental demonstrations. The first concerns a situation where one can implement a given set of measurements and would like to find the best state to demonstrate steering with them. A second case is the reverse: one can produce a given state and wants to find the best set of measurements to demonstrate steering with it. Finally we will also consider a third scenario, where the measurements performed are limited by a certain efficiency and the experimentalist would like to find the best set of measurements and state to demonstrate steering.

Let us start with the first case, where a set of measurements $M_{a|x}$ is fixed and one would like to find the best steerable state $\rho^{\rA\rB}$ for some steering test, quantifier or task (\ie the state which maximally violates a given fixed steering inequality, the state with the biggest steering weight or robustness, or the state with the best local or global randomness). The starting point is to choose a state $\rho^{(1)}$ which is steerable by the measurements $M_{a|x}$.  Notice that any entangled pure state $\ket{\psi}$ is steerable as soon as the measurements $M_{a|x}$ are not jointly measurable\footnote{If $M_{a|x}$ are jointly measurable they will not demonstrate steering on any state.}, so one can simply pick a pure entangled state at random. Then, by using the dual formulation of the quantity of interest, one can follow the see-saw algorithm described in Appendix~\ref{sec: see-saw}. That is, one can iterate a procedure that generates the best steering inequality for the assemblage $\sigma_{a|x}^{(i)}=\tr_\rA[(M_{a|x}\otimes\openone) \rho^{(i)}]$ and then searches for the state $\rho^{(i+1)}$ that violates maximally the previously found steering inequality with the measurements $M_{a|x}$. One can then iterate this procedure until it finds a good pair of state $\rho^{(k)}$ and steering operators $F_{a|x}^{(k)}$.

A special case of the previous approach is to further restrict the class of states one is searching over. For instance, in Ref.~\cite{MGHG14} the authors used this method to find states with positive partial transposition that violate steering inequalities. This demonstrated, for the first time, that bound entangled states can be steerable.

A similar approach can be followed if one has a fixed state $\rho^{\rA\rB}$ and wants to seek a good collection of measurements $M_{a|x}$. In this case there is an additional problem that is to find first a set of measurements that can demonstrate steering with the state $\rho^{\rA\rB}$. Nevertheless, it was shown in Ref.~\cite{SNC14} that random projective measurements are good candidate measurements for demonstrating steering in highly noisy quantum states.

Finally, we discuss the problem of detection efficiency. This is a situation where Alice chooses a set of measurements $M_{a|x}$ with outcomes $a={0,...,o_\rA-1}$, but, in reality,  sometimes no outcome is registered by her measurement apparatus. In order to deal with this situation we will treat the ``no-click" event as an additional outcome. The assemblage that Bob will observe in this case will be given by
\begin{align}\label{eq: assemblage eta}
\assem{\sigma}{a|x}{}(\eta) =
  \begin{cases}
   \eta \sigma_{a|x} &\text{for } a=0,\ldots,o_\rA-1,\\
   (1-\eta)\rho^\rB &\text{for } a=\emptyset.
  \end{cases}
\end{align}
where we have labelled the ``no-click" outcome by $\emptyset$, $\rho^\rB = \sum_a \sigma_{a|x}$ is the reduced state of Bob, and $\eta$ is the probability that Alice's detector clicks. The problem is now to find a set of measurements and a state that can demonstrate steering given a certain detection efficiency $\eta$. First notice that the number of measurements has to be larger than $1/\eta$, otherwise no steering can be demonstrated \cite{BES+12,SC15}. Again, the starting point will be to find an initial assemblage $\sigma_{a|x}^{(1)}$ of the form \eqref{eq: assemblage eta} that demonstrates steering. As shown in Ref.~\cite{SC15}, any set of $m_\rA$ projective measurements can demonstrate steering on any pure entangled state if $\eta>1/m_\rA$ (if $\eta\leq1/m_\rA$ no steering can be demonstrated), so we can choose projective measurements and a pure state at random as a starting point. We now fix $\sigma_{a|x}^{(1)}$, and search for the optimal steering inequality, with operators $F_{a|x}^{(1)}$, that is violated by the assemblage. We then fix the inequality $F_{a|x}^{(1)}$, and search for the assemblage $\sigma_{a|x}^{(2)}$ that maximally violates this inequality and has the form \eqref{eq: assemblage eta}. This process can be repeated until it converges. From the final assemblage $\sigma_{a|x}^{(i)}$ one can finally obtain a physical realisation of it by means of the GHJW construction \cite{G89,HJW93}: Writing $\sum_a \sigma_{a|x}(\eta) = \rho^\rB = \sum_i \lambda_i \ket{i}\bra{i}$, then the Schmidt state,
\begin{equation}
\ket{\psi} = \sum_i \sqrt{\lambda_i}\ket{i}\ket{i}
\end{equation} in combination with the measurements
\begin{align}
M_{a|x} &= \frac{1}{\sqrt{\rho^\rB}} (\sigma_{a|x}(\eta))^T\frac{1}{\sqrt{\rho^\rB}},\nonumber \\
&= \sum_{ij}\frac{1}{\sqrt{\lambda_i\lambda_j}} \ket{i}\bra{j}\sigma_{a|x}(\eta)\ket{i}\bra{j},
\end{align}
which are readily seen to be valid POVMs (positive semidefinite and sum to the identity), always reproduce the desired assemblage.

\subsection{Experimental demonstrations of steering}\label{experiments}

We would finally like to briefly comment on some experimental demonstrations of steering. To the best of our knowledge the first experimental demonstration of steering was realised in the continuous-variable setting, by measuring amplitude and phase quadratures of two entangled optical fields \cite{OPKP92}.

In the discrete case the first demonstration of steering was described in Ref.~\cite{SJWP10} where the authors demonstrate steering by a Werner state that can not violate any Bell inequality for projective measurements.  We would also like to highlight the fist loophole-free steering test of Ref.~\cite{WRS+12}, and the detection loophole-free demonstrations of Refs.~\cite{SGA+12,BES+12}. A demonstration of steering without using steering inequalities has been demonstrated in Ref. \cite{SXY+14}. 

Interesting experimental demonstrations of steering with a single photon shared by two spatial modes were recently reported in Refs.~\cite{FTZ+15,GMM+16}.

The question of one-way steering (\ie states which are steerable from Alice to Bob, but not from Bob to Alice) was experimentally investigated in Ref.~\cite{HES+12} for the case of Gaussian states and Gaussian measurements. More recently it was also demonstrated for qubit-qutrit state with general POVM measurements \cite{WWB+16} and for a two-qubit state when Alice is restricted to apply two dichotomic measurements \cite{SYX+16}.

In the multipartite scenario, Ref.~\cite{CSA+15} provided a demonstration of tripartite steering concerning the definition discussed in Sec.~\ref{Sec: multipartite} and Refs.~\cite{AWT+15,MCS+16} demonstrations of multipartite steering under the definition of Ref.~\cite{HR13} (see Sec.~\ref{sec: other multipartite}).

\section{Conclusions}

{In this article we have reviewed the problem of characterising quantum steering through semidefinite programming. This is a powerful technique that is very suited to the study of many questions related to steering, including its detection, quantification, applications and generalisations. Nevertheless, there are still numerous open questions on this topic still to address, of which we will mention just a few below. 

Concerning the detection of steering, there is no necessary and sufficient conditions to decide whether a quantum state is steerable. Although SDP techniques can be exploited to study this question, when the number of measurement inputs or outputs grow too large the presented SDP techniques usually become computationally too demanding to be practical. As such, it would be interesting to find alternative criteria to determine the steerability of quantum states or assemblages in situations where SDPs become costly. 

On the practical level it is also desirable to recognise real situations where steering plays a major role. This involves asymmetric implementations where due to different laboratory capabilities or to lack of trust some parties are treated differently than other. Moreover, finding experimentally friendly inequalities suitable for real situations involving experimental limitations is an important research direction. In the multipartite setting it would be interesting to find applications where genuine multipartite steering is the key resource. 

On the more foundational perspective, it would also be interesting to understand the extent of post-quantum steering, and whether there are tasks where it provides an advantage over what can be achieved quantum mechanically.

Other interesting open avenues of research concern generalisations of the steering scenario. Previous results have considered cases where the parties receive the inputs encoded in quantum states (rather than as classical information) \cite{2012PhRvL,2013PhRvL}, or when the quantum systems shared between the parties are promised to have certain dimensionality constraints \cite{MGHG14}. We believe that the framework presented here can be useful to attack these and other questions. }

\begin{acknowledgments}
We thank Toni Ac\'in for discussions and M. T. Quintino for comments on the manuscript. This work was supported by the Beatriu de Pin\'os (BP-DGR 2013) and Ram\'on y Cajal fellowship, Spanish MINECO (Severo Ochoa grant SEV-2015-0522), the AXA Chair in Quantum Information Science, and the ERC AdG NLST.
\end{acknowledgments}

\begin{appendix}

\section{Semidefinite Programming Basics}\label{SDP}

In this Appendix, we outline the basic aspects of semidefinite programming that will be necessary for this review. Semidefinite programs are a specific type of convex optimisation problem, with nice analytic properties, and which can be solved efficiently in many cases of interest \cite{VB96}. We will follow closely the presentation of Watrous' lecture notes \cite{W11}, which present SDPs in a form in line with the terminology of quantum information, but further from the presentation found in many other places. The basic SDP optimisation problem is the following:
\begin{align}\label{e:standard primal}
\text{given}& \quad A, \Phi(\cdot), B \nonumber \\
\max_X& \quad \tr[AX] \\
\text{s.t.}& \quad \Phi(X) = B, \nonumber \\
&\quad X \geq 0, \nonumber
\end{align}
where $A^\dagger = A$, $B^\dagger = B$, and $\Phi(\cdot)$ is a hermiticity-preserving linear map. That is, the problem is to maximise the real linear function $\tr[AX]$, over the subset of positive semidefinite operators which satisfy the \emph{linear matrix equality} (LME) constraint $\Phi(X) = B$. The above problem is referred to as the \emph{primal problem} (for reasons that will become clear below). The function $\tr[AX]$ is referred to as the \emph{primal objective} function. The set of operators $X$ which satisfy the constraints ($X \geq 0$, and $\Phi(X) = B$) are said to be \emph{primal feasible}. Finally, the maximum value of the primal objective function over the primal feasible set, denoted $\alpha$ is referred to as the \emph{primal optimal value}. 

To every SDP we can associate the (real scalar) \emph{Lagragrian} functional, by introducing \emph{Lagrange multipliers} for each constraint. For the above SDP, the Lagragrian is
\begin{align}
\mathcal{L} &= \tr[AX] + \tr[Y(B - \Phi(X))] + \tr[ZX], \nonumber \\
&= \tr[(A-\Phi^\dagger(Y) + Z)X)] + \tr[YB],
\end{align}
where $Y^\dagger = Y$, $Z^\dagger = Z$ are the Lagrange multipliers associated to the first and second constraints respectively, and the conjugate map $\Phi^\dagger(\cdot)$ is that which satisfies $\tr[\Phi(X)Y] = \tr[X\Phi^\dagger(Y)]$ for all hermitian $X,Y$. The Lagrangian has the property that $\tr[AX] \leq \mathcal{L}$ for all primal feasible $X$, as long as $Z \geq 0$. That is, the Lagrangian upper bounds $\alpha$. Moreover, $\mathcal{L}$ becomes independent of $X$, and equal to $\mathcal{L} = \tr[YB]$, if the condition $Z = \Phi^\dagger(Y) - A$ is satisfied. Thus, by \emph{minimising} the Lagrangian over the Lagrange multipliers, subject to this condition, we obtain the best possible upper bound on $\alpha$. This is known as the \emph{dual problem}, and is given by
\begin{align}
\text{given}& \quad A, \Phi(\cdot), B \nonumber \\
\min_{Y,Z}& \quad \tr[YB] \\
\text{s.t.}& \quad Z = \Phi^\dagger(Y) -A,  \nonumber \\
&\quad Z \geq 0. \nonumber
\end{align}

More simply, $Y$ and $Z$ are referred to as the \emph{dual variables}. The function $\tr[YB]$ is referred to as the \emph{dual objective} function. The set of operators $Y$, $Z$ that satisfy the LME constraint $Z = \Phi^\dagger(Y) - A$  are said to be \emph{dual feasible}. The minimal value of the dual objective function, denoted $\beta$, is referred to as the \emph{dual optimal value}. Note that -- as often happens -- the above problem can be simplified: the dual variable $Z$ does not appear in the objective function, and the only purpose it serves is to ensure the \emph{linear matrix inequality} (LMI) constraint $\Phi^\dagger(Y) - A \geq 0$. Such variables are referred to as \emph{slack variables} and can always be eliminated to arrive at the simpler form
\begin{align}\label{e:standard dual}
\text{given}& \quad A, \Phi(\cdot), B \nonumber \\
\min_{Y}& \quad \tr[YB] \\
\text{s.t.}& \quad \Phi^\dagger(Y)  \geq A. \nonumber 
\end{align}

By construction, the dual optimal value $\beta$ upper bounds the primal optimal value $\alpha$, which can also easily be seen from $\alpha = \tr[AX^*] \leq \tr[\Phi^\dagger(Y^*) X^*] = \tr[Y^* \Phi(X^*)] = \tr[Y^*B] = \beta$, where $X^*$ and $Y^*$ are the primal and dual variables which achieve the optimum of each problem. This is known as \emph{weak duality}. Much more importantly, it is usually the case that $\alpha = \beta$, which is known as \emph{strong duality}. The condition for strong duality to hold is that either the primal or the dual problem is \emph{strictly feasible} -- that one can find either a positive-definite $X > 0$ such that $\Phi(X) = B$, or a $Y$ such that $\Phi^\dagger(Y) - A > 0$. 

Finally, we note that here we presented only the simplest case of a single optimisation variable, with a single equality constraint. One can easily extend to programs with multiple optimisation variables (e.g. $X_1$, $X_2$, $\ldots$, etc), and to programs with multiple inequality and equality constraints. In all cases, by introducing a Lagrange multiplier for each constraint, and passing to the Lagrangrian, the dual can always be straightforwardly written down.

\section{See-saw algorithm}\label{sec: see-saw}
It is often the case that in the standard primal SDP \eqref{e:standard primal} the operator $B$ appearing in the constraint $\Phi(X) = B$ is itself the output of some linear hermiticity-preserving map $\Lambda(\cdot)$, i.e. $B = \Lambda(C)$, and that the input $C$ is from some feasible set $\{ C | \Psi(C) = D, C \geq 0\}$, where $\Psi(\cdot)$ is another linear hermiticity-preserving map. In this case, the SDP \eqref{e:standard primal} can be thought of as a function of $C$, and there are many instances where one is interested in minimising the primal optimal value over the set of feasible $C$, i.e. one wishes to solve the \emph{min-max problem}
\begin{align}\label{e:standard minmax}
\text{given}& \quad A, \Phi(\cdot), \Lambda(\cdot),\Psi(\cdot),D \nonumber \\
\min_C \max_X& \quad \tr[AX] \\
\text{s.t.}& \quad \Phi(X) = \Lambda(C), && \Psi(C) = D,  \nonumber \\
&\quad X \geq 0, && C \geq 0. \nonumber
\end{align}
This problem is not itself an SDP, however it can be solved heuristically by applying a \emph{see-saw} approach. In particular, although the form \eqref{e:standard minmax} is not suitable to work with directly, when strong duality holds we can equivalently use the dual SDP \eqref{e:standard dual} to express \eqref{e:standard minmax} as
\begin{align}\label{e:standard minmax2}
\text{given}& \quad A, \Phi, \Lambda,\Psi,D \nonumber \\
\min_C \max_Y& \quad \tr[Y\Lambda(C)] \\
\text{s.t.}& \quad \Phi^\dagger(Y) \geq A,  \nonumber \\
&\quad\Psi(C) = D, && C \geq 0. \nonumber
\end{align}
The advantage of this form is that the objective function is now the \emph{bilinear} expression $\tr[Y\Lambda(C)]$ in $Y$ and $C$, and the constraints have decoupled into those which only involve $Y$, and those that only involve $C$. This then suggests the following iterative algorithm: alternatively hold either $Y$ or $C$ fixed, and carry out an SDP optimisation in the other variable. In particular, starting from a feasible $C^{(0)}$ (i.e. $C^{(0)} \geq 0$ and $\Psi(C^{(0)}) = D$, we solve
\begin{align}\label{e:see-saw1}
\max_Y& \quad \tr[Y\Lambda(C^{(0)})]  \\
\text{s.t.}& \quad \Phi^\dagger(Y) \geq A, \nonumber
\end{align}
and denote by $Y^{(0)}$ the minimising $Y$. We then solve 
 \begin{align}\label{e:see-saw2}
\min_C & \quad \tr[Y^{(0)}\Lambda(C)]  \\
\text{s.t.}&\quad\Psi(C) = D, && C \geq 0, \nonumber
\end{align}
and denote by $C^{(1)}$ the minimising $C$. This procedure is then iterated, generating the sequence $\{C^{(i)},Y^{(i)}\}_i$, which is terminated once the sequence converges, up to numerical precision.

We will see that for steering the see-saw approach allows one to solve a number of problems, including finding optimal measurements for a given state in many contexts, and for finding examples of post-quantum steering.

\section{Separability testing}\label{Separability testing}
In \eqref{GMS 1UN}, the last two terms on the right-hand-side both have the property that the each member of the assemblage is an (unnormalised) separable state. Testing for separability is a computationally hard problem \cite{G04}. However, Doherty, Parillo and Spedilieri (DPS) introduced a hierarchy of SDPs which test for membership of quantum states in an outer approximation to the separable set. Moreover, this hierarchy converge in the limit to the separable set itself \cite{DPS02}.

In particular, a state $\rho^{\rA\rB}$ is said to have \emph{$k$-symmetric PPT extension} if there exists a state $\sigma^{\rA\rB_1\cdots \rB_k}$, invariant under interchange of Bob's subsystems, and such that $\rho^{\rA\rB} = \sigma^{\rA \rB_1}$ and $(\sigma^{\rA\rB_1\cdots \rB_k})^{T_{\rB_1\cdots \rB_\ell}}\geq 0$, for $\ell = 1,\ldots, k$, i.e. such that the reduced state of Alice and a single Bob coincides with the state $\rho^{\rA\rB}$, and which is PPT across all bipartitions. DPS showed that a state is separable if and only if it has a $k$-symmetric PPT extension for all $k$. Thus, to possess a $k$-symmetric PPT extension for some fixed $k$ is a necessary condition to be separable. Finally, to check if a state has a $k$-symmetric PPT extension can be solved by SDP.

Using the above, an assemblage is said to have a $k$-symmetric PPT extension if every member possesses an unnormalised $k$-symmetric PPT extension. This can then be decided by the following feasibility SDP:
\begin{align}\label{e:k-ext feas}
\text{given}& \qu \{\assem{\sigma}{a|x}{\rB\rC}\}_{a,x}, \,k \nonumber \\
\text{find}& \qu \{\assem{\pi}{a|x}{\rB\rC_1\cdots C_k}\}_{a,x} \\
\text{s.t.}& \qu \tr_{\rC_2\cdots \rC_k}\left[\assem{\pi}{a|x}{\rB\rC_1\cdots C_k} \right] = \assem{\sigma}{a|x}{\rB\rC} \quad \forall a,x,  \nonumber\\ 
	& \qu \left(\assem{\pi}{a|x}{\rB\rC_1\cdots C_k} \right)^{T_{\rC_1\cdots\rC_\ell}} \geq 0 \qu\forall a,x,\ell\quad \assem{\pi}{a|x}{\rB\rC_1\cdots C_k} \geq 0 \qu \forall a,x, \nonumber\\
	& \qu(\openone_\rB \otimes \Pi_k) \assem{\pi}{a|x}{\rB\rC_1\cdots C_k} (\openone_\rB \otimes \Pi_k)= \assem{\pi}{a|x}{\rB\rC_1\cdots C_k}\qu \forall a,x,\nonumber
\end{align}
where $\Pi_k$ is the projector onto the symmetric subspace of $\mathcal{H}_{\rC_1}\otimes \cdots \otimes \mathcal{H}_{\rC_k}$. If no $k$-symmetric PPT extension can be found, then the assemblage is inconsistent with having only separable members. We will use the notation $\{\assem{\sigma}{a|x}{\rB\rC}\} \in \Sigma^{\rB\rC}_{k-\mathrm{sep}}$ if an assemblage has a $k$-symmetric PPT extension, i.e. if it satisfies the aboves feasibility SDP problem.

\section{Codes}
\label{Sec: codes}
To accompany this review, we will also maintain a small collection of {\sc matlab} code at \cite{code}. This code uses the {\sc cvx}, a modeling framework for disciplined convex programming \cite{cvx,GB08} to implement all of the SDPs, and makes extensive use of the toolbox \emph{QETLAB: A MATLAB Toolbox for Quantum Entanglement} \cite{qetlab}.

As a simple example, let us assume we have a steerable density operator $\rho^{\rA\rB}$ stored in the matlab variable ${\tt rhoAB}$ and a collection of POVMs $M_{a|x}$ stored in the 4-D array ${\tt Max}$, such that ${\tt M(:,:,a,x)}$ equals $M_{a|x}$. Then, to generate an assemblage $\sigma_{a|x}$ we can use the simple function $${\tt sigma = genAssemblage(rhoAB,Max)}$$ to generate a 4-D array ${\tt sigma}$ such that ${\tt sigma(:,:,a,x)}$ is equal to $\sigma_{a|x}$. 

To check that this assemblage satisifes the non-signalling constraints \eqref{nosignaling}, we can call $${\tt NSAssemblage(sigma)}$$ which will return {\tt 1}, to signify that this is the case. Moreover, we can check for normalisation by calling $${\tt NSAssemblage(sigma,1)}$$ which will also return ${\tt 1}$. To certify that $\sigma_{a|x}$ demonstrates steering we can call $${\tt LHSAssemblage(sigma)}$$ which will return ${\tt 0}$. To obtain a steering inequality that certifies this fact we call $${\tt [is\_LHS,siglam,Fax]= LHSAssemblage(sigma)}$$ Here, ${\tt is\_LHS = 0}$ says that the assemblage demonstrates steering, ${\tt siglam = []}$ is the empty array, since there is no LHS model in this instance\footnote{If $\sigma_{a|x}$ was LHS, ${\tt siglam}$ would be a 3-D array such that ${\tt siglam(:,:,\lambda)}$ equals $\sigma_\lambda$.}, and ${\tt Fax}$ is a 4-D array such that ${\tt Fax(:,:,a,x)}$ equals $F_{a|x}$. 

To find the Steering Robustness $\SR(\sigma_{a|x})$, one calls $${\tt SR = steeringRobustness(sigma)}$$ Similarly, to find the local guessing probability of $\sigma_{a|x}$ one calls $${\tt Pg = localSteeringGuessProb(sigma)}$$ 

If one wants to estimate the Steering Robustness of $\rho^{\rA\rB}$ directly, one can call $${\tt SRrho = steeringRobustnessState(rhoAB,Max)}$$ which, using $M_{a|x}$ as the initial set of measurements, will perform the see-saw algorithm given in Appendix~\ref{sec: see-saw} to find the set of measurements (with the same number of inputs and outcomes as $M_{a|x}$) that maximise the steering robustness. One can similarly call $${\tt localSteeringGuessProbState(rhoAB,Max)}$$ to estimate the local guessing probability of $\rho^{\rA\rB}$, starting again with the measurements $M_{a|x}$. 

If one would like to find an example of a qubit-qudit state that has an LHS model for all projective measurements, given an qubit-qudit entanglement witness $W$, stored in the 2-D array ${\tt W}$, one can call $${\tt rhoAB = findPVMLHSStateGivenWitness(W,Max)}$$

The above is a brief demonstration of some of the things which can straightforwardly be achieved using the code provided. A complete list can be found at \cite{code}. 
\end{appendix}


%

\end{document}